\documentclass[12pt,preprint]{aastex}
\usepackage{epsf}
\newcommand{\sss}{$\sim$}
\newcommand{\kms}{km~s$^{-1}$}
\begin{document}
 
\title{EXTREMELY METAL-POOR STARS. VIII. HIGH-RESOLUTION, HIGH-SIGNAL-TO-NOISE ANALYSIS OF FIVE STARS WITH [Fe/H] $\lesssim$ --3.5}

\author{John E. Norris}
 
\affil{Research School of Astronomy \& Astrophysics, The Australian National
University,\\ Mount Stromlo Observatory, Cotter Road, Weston, ACT 2611, Australia;
jen@mso.anu.edu.au}

\author{Sean G. Ryan}
 
\affil{Department of Physics \& Astronomy, The Open University, Walton Hall, Milton Keynes
MK7 6AA, United Kingdom; s.g.ryan@open.ac.uk}

\and

\author{Timothy C. Beers}

\affil{Department of Physics \& Astronomy, Michigan State University, East Lansing, MI 48824; beers@pa.msu.edu
}

\begin{abstract}

High-resolution, high-signal-to-noise ($\langle$S/N$\rangle$ = 85) spectra
have been obtained for five stars -- CD--24$^{\rm o}$17504, CD--38$^{\rm
o}$245, CS~22172--002, CS~22885--096, and CS~22949--037 -- having [Fe/H]
$\lesssim$ --3.5 according to previous lower S/N material.  LTE
model-atmosphere techniques are used to determine [Fe/H] and relative
abundances, or their limits, for some 18 elements, and to constrain more
tightly the early enrichment history of the Galaxy than is possible based on
previous analyses.

We compare our results with high-quality higher-abundance literature data for
other metal-poor stars and with the canonical Galactic chemical enrichment
results of Timmes et al. (1995) and obtain the following basic results: (1)
Large supersolar values of [C/Fe] and [N/Fe], not predicted by the canonical
models, exist at lowest abundance. For C at least, the result is difficult to
attribute to internal mixing effects; (2) We confirm that there is {\it no
upward trend} in [$\alpha$/Fe] as a function of [Fe/H], in contradistinction
to some reports of the behavior of [O/Fe]; (3) The abundances of aluminum,
after correction for non-LTE effects, are in fair accord with theoretical
prediction; (4) We confirm earlier results concerning the Fe-peak elements
that [Cr/Fe] and [Mn/Fe] decrease at lowest abundance, while [Co/Fe]
increases -- behaviors that had not been predicted.  We find, however, that
[Ni/Fe] does not vary with [Fe/H], and at [Fe/H] $\sim$ --3.7, [Ni/Fe] =
+0.08 $\pm$ 0.06.  This result appears to be inconsistent with the supernova
models of Nakamura et al. (1999) that seek to understand the observed
behavior of the Fe-peak elements by varying the position of the model mass
cut relative to the Si-burning regions. (5) The heavy neutron-capture
elements Sr and Ba exhibit a large scatter, with the effect being larger for
Sr than Ba.  The disparate behavior of these two elements has been attributed
to the existence of (at least) two different mechanisms for their production;
(6) For the remarkable object CS~22949--037, we confirm the result of
McWilliam et al. (1995b) that [C/Fe], [Mg/Fe], and [Si/Fe] are supersolar by
$\sim$ 1.0 dex.  Further, we find [N/Fe] = 2.7 $\pm$ 0.4.  None of these
results is understandable within the framework of standard models.  We
discuss them in terms of partial ejection of supernova mantles (Woosley \&
Weaver 1995) and massive (200--500 M${_\odot}$) zero-heavy-element hypernovae
(e.g. Woosley \& Weaver 1982).  The latter model actually predicted
overproduction of N and underproduction of Fe-peak elements; and (7) We use
robust techniques to determine abundance trends as a function of [Fe/H].  In
most cases one sees an apparent upturn in the dispersion of relative
abundance [X/Fe] for [Fe/H] $<$ --2.5.  It remains unclear whether this is a
real effect, or one driven by observational error.  The question needs to be
resolved with a much larger and homogeneous data set, both to improve the
quality of the data and to understand the role of unusual stars such as
CS~22949--037.
\end{abstract}
 
\keywords{stars : nuclear reactions, nucleosynthesis, abundances -- stars :
abundances -- stars : Population II}

\hspace {-1.0mm}{\it Suggested running title:} \hspace*{3mm} FIVE STARS WITH [Fe/H] $\lesssim$ --3.5 

\section{INTRODUCTION}

Two decades ago no stars were known with chemical abundance
[Fe/H]\footnote{[Fe/H] = log~(N$_{\rm Fe}$/N$_{\rm H}$)$_{\rm
star}$--log~(N$_{\rm Fe}$/N$_{\rm H}$)$_{\odot}$} $<$ --3.0 (heavy elements
less that 1/1000 solar) (Bond 1981).  Today, as the result of systematic
searches, in particular the HK survey of Beers, Preston, \& Shectman (1985,
1992), we know of $\sim$ 100 such objects (Beers et al. 1998; Norris 1999).
The most metal-poor stars in these surveys have [Fe/H] = --4.0 (McWilliam et
al. 1995b; Ryan, Norris, \& Beers 1996).

The chemical abundances of objects with [Fe/H] $<$ --3.0 contain clues to
conditions at the earliest times.  As outlined in the previous paper of this
series (Norris, Beers, \& Ryan 2000) they constrain Big Bang Nucleosynthesis
(Ryan, Norris, \& Beers 1999; Ryan et al. 2000), the nature of the first
supernovae (McWilliam et al. 1995b; Ryan et al. 1996; Nakamura et al. 1999),
the manner in which the ejecta from the first generations were incorporated
into subsequent ones (Ryan et al. 1996; Shigeyama \& Tsujimoto 1998;
Tsujimoto \& Shigeyama 1998; Ikuta \& Arimoto 1999; Tsujimoto, Shigeyama \&
Yoshii 1999), and (in special cases) the age of the Galaxy (Cowan et
al. 1999; Cayrel et al. 2001).  Having formed at an epoch corresponding to
redshifts $\gtrsim$~ 4--5, they nicely complement and constrain abundance
results from the more complicated and less well-understood Lyman--$\alpha$
clouds and Damped Lyman--$\alpha$ systems currently studied at redshifts z
$<$ 4.5 (Songaila \& Cowie 1996; Pettini et al. 1997; Prochaska \& Wolfe
1999; Dessauges-Zavadsky 2001).  The field is understandably an active one.

At lowest abundance, the absorption lines upon which analysis is based weaken
dramatically, and one needs to accurately measure features having equivalent
widths $\lesssim$ 10--15 m{\AA}.  For digital spectra, the error associated
with the integration over an absorption line can be written $\sigma_{\rm W}
\simeq$ ($\lambda\sqrt{\rm {n_{pix}}}$)/(R x [S/N]), where R is the resolving
power, S/N is signal-to-noise/pixel, n$_{\rm pix}$ is the number of pixels
over which the integration is performed, and the assumption is made that the
width of the instrumental profile is a fixed number of pixels. If one assumes
further that n$_{\rm pix}$ is constant for different observers, one can
define a figure of merit F = (R x [S/N])/$\lambda$, which should be maximized
for best results.  Table~1 compares F for recent spectroscopic investigations
of stars having [Fe/H] $<$--3.0.  One sees that F varies in the range
85--500~{\AA}$^{-1}$.  This translates, roughly, to 3$\sigma$ errors in line
strength measurement between 35 and 6 m{\AA}, respectively.  To make
definitive progress in this field one thus needs observational material
having F $\gtrsim$ 500.  Perhaps the most comprehensive studies of the past
decade have been those of McWilliam et al. (1995b) and Ryan et al. (1996),
but as may be seen from Table~1 the quality of their data was somewhat
limited, driven by a desire to survey the properties of newly-discovered
objects having [Fe/H] $<$ --3.0.  Their sample sizes, with only 14 and 10
stars, respectively, having [Fe/H] $<$ --3.0, have also limited progress.
One might expect that the high-dispersion spectrographs currently being
deployed on 8--10 m class telescopes will permit the study of larger samples
at higher precision.

In an attempt to provide more accurate abundances at lowest abundance we
decided to obtain high-resolution, high signal-to-noise data for six stars
which, according to earlier analysis, have [Fe/H] $<$ --3.5.  These objects
were chosen with absolutely no bias concerning their relative abundances,
[X/Fe], and within the caveat of small numbers provide a representative
sample of the most metal-poor stars.  Our aim was to obtain accurate relative
abundances with a view to more closely constraining conditions at the
earliest times.  We were particularly interested to establish, where
possible, the lowest abundance anchors to observed relative abundance trends.
We also wished to investigate the question of abundance spreads below [Fe/H]
= --2.5: in the case of [Sr/Fe], for example, there exists a large and
significant spread, but for other elements the situation is presently not
clear.

The results for CS~22876--032, the most metal-poor dwarf in the sample (also
a double-lined spectroscopic binary), have already been reported (Norris et
al. 2000).  Here we present high-resolution (R = 42000), high-signal-to-noise
spectroscopy (S/N \sss~40--100 per 0.04~{\AA} pixel) and elemental abundance
analyses for the remaining five stars.  The objects are CD--24$^{\rm
o}$17504, CD--38$^{\rm o}$245, CS~22172--002, CS~22885--096, and
CS~22949--037.  The reader should see Table~6 of Ryan et al. (1996) for details
of previous analyses.  The mean S/N for the new data set is \sss~85, yielding
F = 830 (as presented in Table~1).

The data and their analysis are reported in \S 2 and \S 3, respectively.  In \S
4 we discuss the implications of our findings.  The results strengthen
earlier conclusions that existing models of supernovae explosions and
Galactic chemical enrichment fail to fully reproduce observed enrichment at the
earliest times.  There is also support for the view that hypernovae (M
$\gtrsim$ 100 M$_{\odot}$ pair-instability supernovae) may have played a
role in that enrichment.

\section{OBSERVATIONAL MATERIAL}

High-resolution spectra were obtained with the University College London
coud\'e\ \'echelle spectrograph on the Anglo-Australian Telescope
during August 1996, 1997, and 1998, and in September 2000.  The instrumental
setup and data reduction techniques were similar to those described by
Norris, Ryan, \& Beers (1996) and Norris et al. (2000), and will not be
discussed here, except to repeat that the material covers the wavelength
range 3700--4700~{\AA} and was obtained with resolving power 42000.  The
numbers of detected photons in the spectra were 10500, 9800, 2500, 8500, and
3300 per 0.04~{\AA} pixel at 4300~{\AA} for CD--24$^{\rm o}$17504,
CD--38$^{\rm o}$245, CS~22172--002, CS~22885--096, and CS~22949--037,
respectively.  (It is perhaps worth noting for completeness that the counts
diminish towards shorter wavelengths, and have decreased to roughly 65\% of
these values at 3900~{\AA}.)  In the case of CS~22172--002 the data were
co-added to our earlier material (Norris et al. 1996) to yield S/N \sss~70 per
0.04~{\AA} pixel.

Examples of the spectra are presented in Figure~1 (which covers the residual
intensity range 0.5--1.0) together with those of the well-studied halo
subgiant HD 140283 ([Fe/H] = --2.5) and giant HD 122563 ([Fe/H] = --2.7)
(from Norris et al. 1996), in the region of the G band of the CH molecule.
The G band is clearly seen in the standard stars, and is surprisingly strong
in CS~22949--037, as was first noted by McWilliam et al. (1995b).  We shall
return to this point in \S 4.

\subsection{Line strength measurements}

Equivalent widths were measured for the lines of 15 ionization species in 13
elements for most of the program stars, together with upper limits for those
of three other important elements using techniques described in Norris et
al. (1996, 2000).  The results are presented in Table~2.  The upper limits to
line strengths listed in the table reflect estimates of the detectability of
lines formed during the interactive line measurement process.  {\it Post
facto} one finds that they are a little larger than formal $\sim$ 3$\sigma$
measurement errors.  (For our spectra $\sigma_{\rm W}$ = 120/(S/N) m{\AA},
leading to 3$\sigma$ errors of 4, 4, 5, 4, and 6 m{\AA} for CD--24$^{\rm
o}$17504, CD--38$^{\rm o}$245, CS~22172--002, CS~22885--096, and
CS~22949--037, respectively).  The corresponding upper limits in Table~2 are
4, 5, 8, 5, and 10 m{\AA}.

Comparisons between the line strengths of the present work and previous
investigations are presented in Figure~2, together with the figure of merit
F, defined above.  The diagram is not meant to be a comparison of the best
data with the best data, but rather to demonstrate the role F plays in
describing the quality of the material.  The agreement is quite satisfactory,
with the scatter in the diagrams being in qualitative agreement with the
values of F.  Quantitatively, the RMS scatter about linear regressions
between the data of different observers were 4, 10, 11, and 17 m{\AA} for
CD--24$^{\rm o}$17504, CD--38$^{\rm o}$245, CS~22885--096, and CS~22949--037,
respectively.  One may compare these with values of $\sigma_{\rm W}$
determined from photon statistics.  We estimate that the corresponding
quadratically added 1$\sigma$ measurement errors are, roughly, 2, 5, 6, and 8
m{\AA}.  While the sense of the agreement is correct, the comparison suggests
that the latter estimates may be somewhat optimistic.

\subsection{Radial Velocities}

Radial velocities were determined with techniques described in earlier papers
of this series (Norris et al. 1996, 1997a, 2000) and Aoki et al. (2000).  The
results are presented in Table~3, together with independent values from the
literature.  We note that column (4) contains the {\it internal} standard
error of the present measurements.  Norris et al. (1996) claim an ``upper
limit to the external standard deviation of a single observation of 1 \kms,"
which should also pertain to the present observations.  Comparison of our
velocities with the independent estimates in Table~3 show no evidence for
variability above the 1 \kms~level in any of the objects for which multiple
observations exist.

\section{CHEMICAL ABUNDANCES}

\subsection{Abundances}

Full details of the techniques of abundance determination have been presented
in Ryan et al. (1996) and Norris et al. (1997a, 2000) and will not be
repeated here, except to draw the reader's attention to the following points.

We assume that line formation takes place under conditions of Local
Thermodynamic Equilibrium (LTE) and adopt the model atmospheres of Bell et
al. (1976) and Bell (1983, private communication).  All calculations are
performed with codes that derive from the original ATLAS formalism (Kurucz
1970; Cottrell \& Norris 1978).  Measured {\it BVRI} colors and estimates of
interstellar reddening, presented in Table~4, form the basis of effective
temperature determination via the calibrations of Bell and Oke (1986), Buser
\& Kurucz (1992), and Bell \& Gustafsson (1978, 1989).  Surface gravity is
determined, in the first instance, by assuming that the star in question lies
on a canonical pre-helium-flash evolutionary track in the color-magnitude
diagram, and refined by the requirement that derived abundances of Fe~I and
Fe~II are identical.  Finally, microturbulent velocities, $\xi$, are
determined by requiring that abundances derived for individual lines should
exhibit no dependence on equivalent width.

The atmospheric parameters for the program stars are presented in columns
(7)--(10) of Table~4, where the column headings should be self explanatory.
Column (9) is our final [Fe/H] and the value adopted for the model atmosphere
calculations.  Relative abundances (together with [Fe/H]), their errors, and
the number of lines used in the analysis are presented in Table~5.  We note
that the Si abundances in the giants are based not only on the equivalent
width of the single unblended 3905.5~\AA\ line (Table~2) but also on spectrum
synthesis of the 4102.9~\AA\ line, which lies in the wing of H$\delta$.

For carbon and nitrogen the abundances are based on spectrum synthesis of the
$\lambda$4323~{\AA} feature of CH and the $\lambda$3883~{\AA} bandhead of CN,
respectively, as described in Norris et al. (1997a).  (In the absence of
information on the abundance of oxygen we assume [O/Fe] = 0.6, and note that
for the program stars the results are insensitive to variations of
$\Delta$[O/Fe] = $\pm$0.4.)  A comparison of observed and synthetic spectra is
presented in Figure~3 for four of the program stars, as well as for the more
metal-rich Population II stars HD~122563 and HD~140283\footnote {Full details
for the standards may be found in Norris et al. (1996, 1997a) and Ryan et
al. (1996).}.  Carbon is detected in three of the five program stars, and for
CS~22885--096 and CS~22949--037 the present results confirm earlier values
reported by McWilliam et al. (1995b), albeit with somewhat higher accuracy.

Nitrogen is detected in only CS~22949--037.  The claimed overabundance,
[N/Fe] = 2.7 $\pm$ 0.4, is both astounding and, at first sight, almost
unbelievable. In an effort to convince ourselves that the result was not an
artifact of the measurement procedure (in particular faulty cosmic ray
removal) we re-examined the observational material in some detail.  In the
initial analysis, cosmic rays had been removed using the BCLEAN and CLEAN
utilities of the reduction package FIGARO.  In the re-analysis they were
removed in two different ways. On each of the four nights during which the
data were obtained, median and 2$\sigma$ clipped average images were obtained
and then added to form the final spectra.  In both cases the essential result
remained unchanged.  Further, the CN bandhead is seen in the data subsets on
individual nights.  As a final check, the reduction was repeated by a
different author using the IRAF reduction package.  The same result was
obtained.

After the present analysis was completed, a service observation of
CS~22949--037 was obtained with the William Herschel Telescope/Utrecht
\'echelle spectrograph combination.  It provides confirmation of the
existence of an absorption feature at 3883~\AA\ from an independent dataset,
albeit at much lower S/N = 14.

\subsection{Random errors}  

The abundance errors for atomic species deserve comment.  Random errors arise
from those associated with line-strength measurements and with the
determination of atmospheric parameters (uncertainties in colors, etc.),
while systematic ones result from the (color, effective
temperature)-calibrations and the analysis procedure itself (assumption of
LTE, choice of model atmosphere, etc.).  The errors listed in Table~5 address
only random errors: they represent the quadratic addition of the standard
error in abundance (the error in the mean abundance) derived from individual
lines\footnote{In order to obtain a conservative estimate, the standard error
for a given species is taken as the larger of the values determined from the
available lines of that species and $\langle$s.d.$\rangle$/$\sqrt{{\rm n}}$,
where $\langle$s.d.$\rangle$ is the mean standard deviation over all elements
for the star and n is the number of lines for the species in question.} and
errors resulting from atmospheric parameter uncertainties $\Delta$T$_{\rm
eff}$ = 100~K, $\Delta$log~$g$ = 0.3, and $\Delta\xi$ = 0.25 km s$^{-1}$,
which we believe to be appropriate for the present investigation.

\subsection{Systematic errors}

Systematic errors are more difficult to assess. We have discussed elsewhere
(Ryan et al. 1999) differences that arise between alternative plausible
choices of effective temperature scales.  For example, Alonso, Arribas, \&
Martinez-Roger (1996) have determined temperatures using the Infrared Flux
Method (IRFM) that are generally higher than those from {\it BVRI} colors and
model atmospheres.  In order to quantify the effect, we assessed the impact
on our analysis had we adopted IRFM-related temperatures.  Due to the
$\sim$1000~K difference between CD--24$^{\rm o}$17504 and the giants, we
address these two groupings separately.

The turnoff/subgiant star CD--24$^{\rm o}$17504 has an IRFM T$_{\rm eff} =
6375$~K (Alonso, private communication to Primas et al. 2000), some 300~K
hotter than adopted here. To assess the effect of using the IRFM scale, we
repeated our analysis using that higher temperature.  The resulting
atmospheric parameters were T$_{\rm eff}$/log~$g$/[Fe/H]/$\xi$ =
6375/4.2/$-$3.15/1.4, indicating a surface gravity higher by 0.6~dex, and
[Fe/H] higher by 0.22~dex.  (Inferred \ion{Fe}{1} abundances in Population~II
turnoff stars increase by $\sim~0.07$--0.08~dex per 100~K.)  Columns (2) and
(3) of Table~6 present abundances determined in the present work and those
obtained by using IRFM based atmospheric parameters, respectively, while
column (4) contains relative abundance differences in the form $\Delta$[X/Fe]
= [X/Fe]$_{\rm IRFM}$ $-$ [X/Fe]$_{\rm This~work}$. All differences are
relatively small: typical (RMS) differences are 0.06~dex for neutral lines
and 0.12~dex for ionized lines (excluding \ion{Fe}{2}, the agreement of which
was forced in determining log~$g$). These are comparable to the 1$\sigma$
random errors already listed in Table~5, whose RMS values for neutral and
ionized lines are 0.10 and 0.14~dex, respectively, in this star.

For the four giants of the present study, IRFM temperatures are unavailable,
but it is possible to transform the BVRI colors to the IRFM scale using the
color-metallicity-temperature calibrations provided by Alonso, Arribas, \&
Martinez-Roger (1999). Using the Johnson-Cousins photometry from our Table~4,
Cousins-to-Johnson transformations from Bessell (1979), and Johnson-to-IRFM
calibration equations from Alonso et al.  Table~2, we find the temperatures
given in Table~7. The heading of each IRFM column includes the equation
number of the calibration.  It was necessary to extrapolate outside the
stated range of validity of the Alonso et al. calibrations (see their
Table~3) because of the dearth of calibrating stars with [Fe/H] $<$ $-3$;
although such extrapolations can be misleading, IRFM temperatures are
nevertheless being adopted by many other workers in this field, and a
comparison of the two scales may be instructive.  The entries based on $V-R$
and $R-I$ are flagged as uncertain because the calibrations show greater
sensitivity to metallicity between [Fe/H] = --3 and [Fe/H] = --2 than they do
between [Fe/H] = --2 and [Fe/H] = --1, which is unlikely to be correct; see
Alonso et al. (1999) Table~6 and the discussion of this issue by Ryan et
al. (1999), especially their Fig.~5.  The final column of Table~7 gives for
each star the difference between the mean of the IRFM-scale temperatures and
the temperature we have adopted in Table~4.  The differences are small in all
cases except one, $\Delta$T$_{\rm eff}$ = 103~K for CS~22949--037, but even
that difference is smaller than the formal 1$\sigma$ errors associated with
the Johnson-to-IRFM calibrations of 96--167~K (see final row of Table~7,
transcribed from Alonso et al. Table~2).  The comparison suggests that even
if we had adopted IRFM-scale temperatures for the giants, the changes would
have been minor.  Columns (5)--(7) of Table~6 show the abundance changes that
result for CS~22949--037 if we adopt the higher temperature, which leads to
atmospheric parameters T$_{\rm eff}$/log~$g$/[Fe/H]/$\xi$ =
5000/1.9/$-$3.72/2.0.  The largest change in [X/Fe] is 0.07~dex, and for most
species it is considerably smaller. (We note for completeness that adoption
of the hotter model would lead to higher relative abundances of carbon and
nitrogen by 0.05~dex.)

As discussed by Ryan et al. (1996, \S 3.2.1) the adoption of different model
atmospheres will also lead to small systematic differences.  They found
$\Delta$[Fe/H] $\sim$ 0.10 and 0.15 for dwarfs and giants, respectively,
between analyses using Kurucz (1993) models and those of Bell and coworkers
used here.  The origin of this effect was attributed to differences of order
200~K in the temperature in the line forming regions, arising from the higher
convective energy transport of the Kurucz models.

During the analysis of the {\it giants}, a curious dependence of derived
abundance on excitation potential $\chi$ was noticed for the neutral iron
lines.  (It is not clear whether other atomic species are affected; neutral
iron has the largest number of lines, so any abnormal effects are clearer.)
\ion{Fe}{1} lines in the range $\chi$ = 1.4--3.3~eV show no dependence on
$\chi$, but the lower excitation lines give an abundance higher by
0.2--0.3~dex at $\chi$ = 0. Note that no such trend is seen either for the
dwarf, CD$-$24$^\circ$17504, or in our analysis of the higher-metallicity
halo giant HD~122563 ([Fe/H] = $-2.68$), for which we also have high-quality
data (Ryan et al. 1996).  Although the microturbulence $\xi$ can in principle
affect $\chi$-residuals (since on average lower-excitation lines tend to give
rise to stronger lines), tests showed that it was not possible by adjusting
$\xi$ to remove the $\chi$ trend and simultaneously get a sensible [Fe/H] versus
equivalent width relationship.

We considered the possibility that errors in the temperature gradients of our
adopted one-dimensional models might be too shallow, i.e. the outer layers
too hot compared with the result for three-dimensional models found by
Asplund et al. (1999). Low-$\chi$ lines having (on average) higher line
strength and hence forming (on average) further out would then be computed
too weak and higher abundances would be derived. However, the exclusion of
stronger lines had little effect on the pattern, and the fact that the dwarf
was not affected, whereas Asplund et al's analysis was for dwarfs, made us
doubt this possibility.

McWilliam et al. (1995b) reported $\chi$ trends when they used Bell \&
Gustafsson (1978, 1989) T$_{\rm eff}$ scales, so they used those calculations
only to trace the metallicity sensitivity, and tied the colors to a
solar-metallicity calibration of McWilliam (1990). They derive temperatures
approximately 100~K cooler. However, an arbitrary decrease of 100~K in the
effective temperatures of our models changes only the overall slope of the
$\chi$-trends (and to the degree expected from stellar atmosphere theory); it
does not alter the distortion of low- versus high-excitation lines. We can only
surmise that McWilliam et al. found a simpler trend.

We tried replacing the modified van der Waal's damping values used by Ryan et
al. 1996, \S 3.3) with those from Anstee \& O'Mara (1995), Barklem \& O'Mara
(1997), and Barklem, O'Mara, \& Ross (1998), but this had very little effect
($\le 0.03$~dex) on the abundances and none at all on the
$\chi$-trends. (Very metal-poor stars have few high-excitation-potential
lines, so are much less affected by the $\chi$-dependent damping errors
discussed by Ryan (1998).)

An alternative is that non-LTE effects are present, possibly consistent with the 
peculiarity showing up in the giants but not the dwarf. One possible effect of 
relying on an LTE analysis is that \ion{Fe}{2}/\ion{Fe}{1} ratios produce the 
wrong gravity (see below). We examined the \ion{Fe}{1}/\ion{Fe}{2} and 
\ion{Ti}{1}/\ion{Ti}{2} ratios to see what gravities would be inferred. The 
differences were random, and suggested an internal uncertainty 
$\Delta$~log~g $\la$~0.3 dex, entirely consistent with the observational 
uncertainties in [\ion{Fe}{2}/\ion{Fe}{1}] of 0.07--0.13~dex. 
(A 0.10~dex error in the abundance ratio leads to a 0.33~dex error in the 
gravity.) Trying giants with gravities arbitrarily 0.8~dex higher improved the 
picture only slightly, leaving us unconvinced that this was the correct 
(or even a viable) solution.

The assumptions of LTE, one-dimensional models, and the treatment of
convection in terms of the mixing-length approximation all have the potential
to seriously compromise the results presented here.  Non-LTE effects can be
important, as demonstrated by Baum\"{u}ller and Gehren (1997) for the
resonance lines of Al I (see \S 4.1.4 below).  For other elements, however,
there exist conflicting results: for iron, e.g., Th\'evenin \& Idiart
(1999) and Gratton et al. (1999) reported quite different conclusions. The
former found a large degree of over-ionization in metal-poor stars, which they
attributed to the lower line opacity and hence higher (more important) UV
radiative flux in low-metallicity atmospheres. They suggested two
consequences: (1) spectroscopically-determined log~$g$ values based on LTE
analyses will be too low, by up to 0.4--0.6~dex for dwarfs at [Fe/H] $\sim
-3$, and (2) abundances derived from the minority ionization state
(\ion{Fe}{1}) will be too low by $\sim$ 0.3~dex at [Fe/H] $\sim -$3.  A
comparison of trigonometric and spectroscopic log~$g$ determinations by
Allende Prieto et al. (1999) showed a similar metallicity dependence, but
their analysis also emphasised that the magnitude of the correction varied
considerably from one study to the next (e.g. their Figures~7 and 10), such
that corrections derived in one study could not necessarily be applied to
another.  Gratton et al. (1999), in contradistinction to the other two
works, found no strong effect in dwarf atmospheres.

It has also become clear that realistic three-dimensional modeling of
metal-poor stars, and the consequent rigorous treatment of stellar
convection, may significantly modify the abundances presented here.
Preliminary results of Asplund et al. (1999) for stars having [Fe/H] $\sim$
$-2.5$ show that modifications to LTE 1D model abundances of order 0.2 dex
may be necessary, though they also found (Asplund \& Nordlund 2000) that for
\ion{Li}{1} at least, non-LTE 3D computations gave very similar abundances to
LTE 1D calculations.

Clearly both non-LTE computations and 3D modeling are rapidly developing fields
whose ultimate implications have yet to become clear. The reader should bear
these caveats in mind when reading the following section.

\section{DISCUSSION}

\subsection{Relative Abundances, [X/Fe], as a Function of [Fe/H]}

The present results permit us to re-visit relative abundance trends as a
function of [Fe/H], following Ryan et al. (1996) and Norris et al. (1997a).
Figures~4--9 present [X/Fe] versus [Fe/H] based on the results in
Table~5 and those from the literature.

\subsubsection{Carbon \& Nitrogen}

Figure~4 presents [C/Fe] as a function of [Fe/H], where the results from
Table~5 are supplemented by data from McWilliam et al. (1995b), Norris et
al. (1997a), Ryan et al. (1991), and Tomkin et al. (1992, 1995).  Stars with
[C/Fe] $>$ +1.5 are not considered in the present discussion.  (These include
objects with very large over-abundances of s-process heavy elements (Norris
et al. 1997a; Aoki et al. 2000; Hill et al. 2000) that are believed to have
resulted from mass transfer across a binary system, together with the
enigmatic CS~22957--027 ([Fe/H] = --3.4, [C/Fe] = +2., but ``normal'' heavy
neutron-capture element abundance; Norris, Ryan, \& Beers 1997b, Bonifacio et
al. 1997).  While the s-process enriched objects tell us about
nucleosynthesis in intermediate mass stars, it remains to be seen what the
relationship is between CS~22957--027 and the stars studied here.)  Also
shown in the figure is the standard stellar evolution, Galactic chemical
enrichment (GCE) prediction of Timmes et al. (1995) (without modification of
their computed Fe yield).

The data in the present paper strengthen the conclusion that there is a large
spread in carbon at lowest abundance. At [Fe/H] $<$ --3.0, the range in
[C/Fe] is of order 1 dex.  It is important to note that this large spread is
not an observational selection effect: none of the stars with [Fe/H] $<$
--3.0 in Figure~4 was chosen for study with any knowledge relating to its
carbon abundance. (Rossi, Beers, \& Sneden (1999) have discussed large
numbers of metal-poor, C-rich stars, but for a C-enhanced selected
sub-sample, drawn {\it post facto} from the HK medium-resolution follow-up
spectroscopy campaign.)  Part of the spread (for giants) may result from the
internal mixing of CNO processed material into outer layers during
giant-branch evolution as has been invoked to explain carbon depletions
observed in the most metal-poor globular clusters (Kraft et al. 1982; Langer
et al. 1986).  Such an explanation is not inconsistent with the fact that
CD--38$^{\rm o}$245 (Fe/H] = --3.98, [C/Fe] $<$ 0.00) is nitrogen rich
([N/Fe] = 1.7, with a possible zero-point error of 0.6 dex, Bessell \& Norris
1984)\footnote{We note for completeness that spectrum synthesis calculations
(see \S 3.1) for CD--38$^{\rm o}$245 with [C/Fe] = 0.0 and [N/Fe] = 1.7
predict an undetectable violet CN band in the present spectra, consistent
with the observed spectrum shown in Figure 3.}.  If such mixing has indeed
occurred, any giant that now has [C/Fe] $>$ 0.0 could have originated with an
even larger carbon abundance. The more important point here, however, is that
whatever the role of mixing, real and large supersolar [C/Fe] values exist at
lowest metallicity.  Furthermore, as may be seen in Figure~4, such
overabundances are not predicted by canonical GCE models, suggesting that
simulations of the most metal-poor supernovae or mass-loss during late stages
of massive-star evolution are incomplete.  We shall return to this point in
\S4.2.

As noted in \S 3.1, the value of [N/Fe] = +2.7 for CS~22949--037 is
surprisingly large.  If, in the absence of any knowledge concerning the
abundance of oxygen, one were to suppose that the overabundance of nitrogen
arose from internal mixing of equilibrium CN-cycle processed material into
the star's outer layers as discussed above, one is led to an initial carbon
abundance [C/Fe] = +2.1.  The alternative is that both carbon and nitrogen
were enormously enhanced in the ejecta of the object which enriched the
material from which CS~22949--037 formed.  We defer consideration of this
possibility to \S 4.2, following discussion of the abundances of the other
elements.

\subsubsection{Heavier Elements}

For elements heavier than carbon we use comparison material based on
observations similar in quality to those in Table~2, by only accepting
abundances derived from data having figure of merit F $>$ 300 (F is defined
in \S 1).  (This choice of the limit is of course arbitrary, since the
accuracy of abundances will be a function of the number of lines and the
strength of those lines for a given element. We seek here only to apply the
coarsest of cuts to literature sources.)  Where possible we have corrected
the published values for differences between the literature solar abundances
and those adopted here (Table~5, column (3)).  The literature
sources\footnote{We have excluded carbon-rich objects such as those studied
by Norris et al. (1997a,b), and have not sought to include results from works
which investigate only one or two astrophysically interesting but sometimes
unusual objects.  The fascinating r-process-enhanced star CS~22892--052 has
thus not been included in the comparisons.} are Gilroy et
al. (1988)\footnote{For only the heavy neutron-capture elements, which
represent the thrust of that work.}, Gratton (1989), Gratton et al. (1987,
1988, 1991, 1994), Ryan et al. (1991, 1996), Nissen \& Schuster (1997),
Carney et al. (1997), Stephens (1999), and Fulbright (2000).  (We discuss in
the Appendix a modification we have made to a critical literature $gf$ value
for Co I.)

The comparisons are presented in Figures~5--9\footnote {The data used in these
plots are available via the web address {\tt
http://physics.open.ac.uk/$\sim$sgryan/nrb01\_apj\_data.txt}.}  . As in Ryan et
al. (1996) the left panels of the figures permit one to examine the origin of
the data. (Error bars are attached to stars from the present study, and to
CS~22876--032 ([Fe/H] = --3.71, Norris et al. 2000; open circle), while
results for objects having more than one analysis are connected.  For Ti we
plot the results for Ti II, which differ insignificantly from the average of
Ti I and Ti II (weighted by the number of lines involved.)) The right panel
presents the data anonymously, together with abundance trends determined with
robust statistical tools.  The thin lines represent {\it loess} regression
lines described by Cleveland (1994), and determined as follows.  First,
average values of each abundance ratio were obtained if results were
available from different authors.  Next, we obtained three vectors -- the
central loess line (CLL), the lower loess line (LLL), and the upper loess
line (ULL), as a function of the [Fe/H] values.  The CLL is defined as the
loess line obtained when all the data are considered, and provides our best
estimate of the general trend of the elemental ratios at a given [Fe/H] --
this line replaces the midmean trend line considered previously by Ryan et
al. (1996).  Next, residuals about the CLL were obtained, and separated into
those above (positive residuals) and below (negative residuals) this line.
The LLL is defined as the loess line for the negative residuals as a function
of [Fe/H] -- this replaces the lower semi-midmean considered by Ryan et
al. (1996).  The ULL is defined as the loess line for the positive residuals
as a function of [Fe/H] -- this replaces the upper semi-midmean considered by
Ryan et al. (1996).  If the data are scattered about the CLL according to a
normal distribution, the LLL and ULL are estimates of the true quartiles.
The loess summary lines have an advantage over the midmeans and semi-midmeans
used by Ryan et al., in that one is not forced to arbitrarily bin the data in
order to preserve resolution while suppressing noise.  The loess lines remain
sensitive to local variations without being unduly influenced by outliers.
Furthermore, they are able to better handle the endpoints of the data sets.
In each subpanel the CLL is flanked by the ULL and CLL.  Also shown in the
figures as thicker dashed lines are the GCE predictions of Timmes et
al. (1995).

\subsubsection{The $\alpha$-Elements: Mg, Si, Ca, and Ti.}

The trends in Figure~5 collectively show that there is no strong change of
[$\alpha$/Fe] with [Fe/H] in the range --4 to --1, consistent with previous
investigations.  This is in marked contrast with recent reports that [O/Fe]
increases monotonically as [Fe/H] decreases, with d[O/Fe]/d[Fe/H] $\sim$
--0.33 in the range --3.0 $<$ [Fe/H] $<$ 0.0 (Israelian, Garcia L\'opez, \&
Rebolo 1998; Boesgaard et al. 1999).  Elements lighter than calcium are
chiefly produced during hydrostatic burning in stars, and the GCE models of
Timmes et al. (1995) show little dependence of [O, $\alpha$/Fe] on [Fe/H]
under the assumption that hydrostatically burned regions are expelled during
the supernova phase.  Insofar as theory requires a coupling between the
behavior of O and $\alpha$-elements, the present results do not offer support
for the claimed upward trend of [O/Fe] with decreasing [Fe/H].  That said, we
refer the reader to Israelian et al. (2001) for the case that ``there is a
range of parameters in the calculations of nucleosynthesis yields from
massive stars at low metallicities that can accommodate (an increase of
[O/Fe] with decreasing [Fe/H])"\footnote {We note for completeness that the
case for increasing [O/Fe] is based on standard analysis of near-ultraviolet
OH lines and near-infrared O I lines, and is in conflict with the
non-increasing behavior derived from the [O I] 6300~{\AA} line
(e.g. Fulbright \& Kraft 1999).  That said, the reader should also see Cayrel
et al. (2000) for an opposing view.  Asplund et al. (1999) show that the
outer layers of 3D metal-poor model atmospheres are cooler than 1D model ones
by several hundred K.  Preliminary investigations (Asplund et al. 2000)
suggest that oxygen abundance measurements utilizing the OH molecule may be
particularly sensitive to this difference and could severely overestimate the
true elemental abundance.  If this is in fact the case, O abundance
measurements based on LTE analysis of atomic features may be more reliable,
and the proposed increase of [O/Fe] with declining [Fe/H] may cease to
conflict with the observed [$\alpha$/Fe] behavior (modulo resolution of the
conflicting results for visible and near-IR lines).}.  We note also, as
emphasized to us by the referee, that decoupling between O and the
$\alpha$-elements exist in some of the supernova models reported by Woosley \&
Weaver (1995).

Excluding CS~22949--037, the relative $\alpha$-element abundances
of the stars studied here appear to be quite similar, except for Si, where a
large spread is evident.  Given that the Si values are not as well-determined
as those for the other elements, we believe it would be premature to
interpret the large scatter of this element as originating from other than
error of measurement, and we shall thus not discuss the apparent Si spread
further.

Concerning CS~22949--037, we confirm the result first reported by McWilliam
et al. (1995b) that [$\alpha$/Fe] is larger in this star than in other extremely
metal-poor stars.  We also find that the size of the difference appears to
decrease with increasing atomic number within the $\alpha$-element class: if we
compare CS~22949--037 with (as a group) the other four stars studied here we
obtain $\Delta$[Mg/H] = 0.77 $\pm$ 0.14, $\Delta$[Si/H] = 0.75 $\pm$ 0.31,
$\Delta$[Ca/H] = 0.25 $\pm$ 0.21, and $\Delta$[Ti/H] = 0.07 $\pm$ 0.17.  The
corresponding values from the results of McWilliam et al. for CS~22949--037
relative to their mean sample values are 0.80 $\pm$ 0.16, 0.35 $\pm$ 0.33,
0.46 $\pm$ 0.15, and 0.22 $\pm$ 0.15, respectively.  The agreement is
excellent.

One may re-visit the discussion of McWilliam et al., who noted that these
results are suggestive of only partial ejection of the stellar mantle during
the supernovae explosion(s) that enriched the material from which
CS~22949--037 formed.  Woosley and Weaver (1995) report supernovae
simulations with relatively low ejection velocities in which no Fe is
ejected, but with expulsion of some of the lighter elements.  Consider, in
particular, their model Z35B for a zero heavy element, 35 M$_{\odot}$,
intermediate-energy explosion in which they report production
factors\footnote{The production factor is the ratio of an isotope's mass
fraction in the total ejecta divided by its corresponding mass fraction in
the Sun.} of 4.2, 7.2, 3.3. 0.09, 0.00, 0.00, and 0.00 for $^{12}$C,
$^{16}$O, $^{24}$Mg, $^{28}$Si, $^{40}$Ca, $^{48}$Ti, and $^{56}$Fe,
respectively.  The atomic-mass dependence of the yield of this model is
broadly able to explain the trends seen in CS~22949--037, though the
particular model quoted produces little Si.  A slightly lower location,
however, of the mass-cut presumably might rectify this mismatch, as $^{40}$Ca
and $^{56}$Ni (the parent of $^{56}$Fe) are formed deeper than much of the
$^{28}$Si and all of the $^{24}$Mg.  In principle it might be possible to
eject larger amounts of Mg and Si while ejecting little Ca and iron-peak
elements.  (See likewise the models of Umeda, Nomoto, \& Nakamura (2000).)  We
shall discuss this question further in \S 4.2.1.  

Nucleosynthesis models currently neglect mixing of deeper material into the
outer layers, although at least some observations of SN~II (Fassia et
al. 1998; Fassia \& Meikle 1999) and their ejecta (Travaglio et al. 1999)
seem to require it.  Such mixing may provide an alternative means of
producing partial iron-peak element enrichment of the ejected Mg--Si envelope.

\subsubsection{Aluminum}

The results for [Al/Fe] and [Al/Mg] are presented in Figure~6.  They are
based on only the Al I 3961.5~{\AA} line, given the possible contamination of
Al I 3944.0~{\AA} by CH, as first suggested by Arpigny \& Magain (1983), and
the support for this claim by our finding that the line is broader than
expected in C-enhanced stars (Ryan et al. 1996, \S 3.4.4).

For stars of the lowest metallicity, Al abundances are necessarily based on
the resonance lines used here, since other lines of the species lie below the
detection limit.  This unfortunately leads to large systematic errors.
Baum\"{u}ller \& Gehren (1997) have demonstrated that for dwarfs with [Fe/H]
$\sim$ --3.0, LTE analysis of Al I 3961.5~{\AA} leads to an error
$\Delta$[Al/H] = --0.65.  For [Fe/H] $>$ --2.0, in contrast, use of the Al I
lines near 6697~{\AA} leads to errors of only $\sim$ --0.15.  It remains to
be seen how large the non-LTE corrections are for giants.  We note that for
[Fe/H] $<$ --3.0 our analysis yields essentially the same behavior of [Al/Fe]
for both giants and dwarfs.

We also comment on the distribution of the data in the ([Al/Fe], [Fe/H])- and
([Al/Mg], [Fe/H])-planes for [Fe/H] $>$ --2.0, where one sees that the scatter
in [Al/Mg] is considerably smaller than that in [Al/Fe].  Since Al and Mg are
synthesized in the same region of a star, while Fe is produced in deeper
layers, one might not be too surprised to find the stronger correlation between
Al and Mg.  The trend appears to reverse for [Fe/H] $<$ --2.0, but is based
on fewer data, and may be compromised by gravity dependent non-LTE effects.
It will be interesting to see if this behavior proves spurious as more,
higher-quality, material is obtained.

Given the role of non-LTE effects, it is difficult (and perhaps unwise) to
attempt to compare the observed values with the predictions of the GCE models
of Timmes et al. (1995).  In a zeroth-order approach, however, we seek to use
the results of Baum\"{u}ller \& Gehren (1997) to correct the abundances
presented here.  Specifically, we assume that LTE abundances based on the Al
I 3961.5~{\AA} resonance line should be increased by 0.65 dex, while those
derived from the lines near 6697~{\AA} should be corrected by +0.15 dex.  In
the absence of conflicting information we apply the same corrections to both
giants and dwarfs.  Figure~7 then presents the comparison of (corrected)
observation and theory for [Al/Mg] versus [Fe/H].  At lowest abundance the
agreement is quite satisfactory.  That said, we refer the reader to the work
of Baum\"{u}ller \& Gehren (1997, \S 4.2) for interesting differences between
the theoretical predictions and non-LTE abundances at [Fe/H] $\sim$ --1.0.

The difference in abundance between CS~22949--037 and the mean of the other
four stars of the present study is $\Delta$[Al/H] = 0.43 $\pm$ 0.32, which is
not statistically significant given the errors.

\subsubsection{The Iron-peak Elements} 

Figure~8 presents the observed behavior of the [Fe-peak/Fe] as a function of
[Fe/H].  The downturn of [Cr/Fe] and [Mn/Fe] and the upturn [Co/Fe] below
[Fe/H] = --2.5 were first reported by McWilliam et al. (1995b), and confirmed
by Ryan et al. (1996).  The behavior of Co in the currently more-restricted
and refined (see Appendix) data set bears comment.  In Figure~8 one sees a
relatively rapid rise to supersolar values at [Fe/H] $\sim$ --2.5.  It will
be interesting to see if larger samples confirm the effect.

The Cr, Mn, and Co trends had not been predicted by theory, but subsequently
Nakamura et al. (1999) suggested that the inadequacy might result, at least
in part, from inadequate knowledge of the mass cut above which material was
ejected.  By moving the cut relative to the regions of complete and
incomplete Si-burning, they were able to reproduce the above trends.
Nakamura et al. (2000) reported that the behavior of Cr, Mn, and Co could
also be produced by explosive nucleosynthesis in hypernovae, defined by them
as objects having very large explosion energies ($\gtrsim$ 10$^{52}$ ergs),
which have more extended regions of complete and incomplete Si-burning.

The solution of Nakamura et al. (1999) is, however, only a partial one, since
they predicted that [Ni/Fe] would exhibit the same behavior as [Co/Fe], which
was not confirmed by existing data.  The present material more strongly
constrains the problem than previously possible.  For the five stars reported
here, we find $\langle$[Ni/Fe]$\rangle$ = +0.08 $\pm$ 0.06, which differs
from the value $\langle$[Co/Fe]$\rangle$ =  +0.57 $\pm$ 0.02 at a high level of
significance.  That is, there is no upturn in [Ni/Fe] at the lowest
abundances.

We note in concluding this section that the work of Nakamura et al. (2000) is
interesting for another reason.  The higher energy of their hypernovae models
has the suggestive beneficial outcome that it provides a possible solution,
{\it for the first time}, to the long-standing problem of why Ti is
overabundant in halo stars.  The authors report that the greater radial
extent over which Si-burning occurs in high-energy models necessarily
incorporates lower density zones, which allows for increased $\alpha$-rich
freeze-out production of $^{48}$Ti.  Previous classical supernova
nucleosynthesis models always yielded [Ti/Fe] $\simeq$ 0 (see e.g. our Figure
5).

\subsubsection{Heavy Neutron-Capture Elements}

Results for [Sr/Fe] and [Ba/Fe] are presented in Figure~9.  As first noted
by Ryan et al. (1991) there exists a large spread in [Sr/Fe] below [Fe/H] =
--3.0.  Our belief in the indisputable reality of the effect was discussed at
some length by Ryan et al. (1996) which we shall not repeat here, and to
which we refer the reader.  The spread is not as large for [Ba/Fe] in the
present restricted data set, and the bimodal distribution of [Ba/Fe]
discussed by Ryan et al. (1996) for [Fe/H] $<$ --2.5 was eliminated by
McWilliam's (1998) revised analysis of the McWilliam et al. (1995a) data
taking account of hyperfine structure of Ba.  The present results confirm our
earlier one that the trends for [Sr/Fe] and [Ba/Fe] are quite different, with
a larger scatter existing at lowest abundance for [Sr/Fe].

Possible causes of this difference have been considered by McWilliam (1998),
Blake et al. (2001) and Ryan et al. (2001), who suggested the need for two
processes to explain the data.  The first is the basic r-process which is
clearly necessary to explain the well-established pattern of abundances at
large atomic number in stars with [Fe/H] $<$ --2.5 (e.g. Sneden \&
Parthasarathy 1983), while the second is an ill-defined one which favors
production of heavy neutron-capture elements at lower atomic number.  Studies
of additional metal-poor stars that might illuminate this discussion are
presently underway.

\subsubsection{Abundance Scatter as a Function of [Fe/H]}

In order to quantify the abundance scatter in these diagrams we compute the
scale\footnote {The scale matches the dispersion for a normal distribution.}
of the data for each elemental ratio, making use of the CLL obtained above,
and consider the complete set of residuals in the ordinate of each data point
about the trend.  In Table~8 we summarize robust estimates of the scale of
these residuals over several ranges in [Fe/H], using the biweight estimator
of scale, $S_{BI}$, described by Beers, Flynn, \& Gebhardt (1990).  The first
column of the table lists the abundance ranges considered.  In setting these
ranges, we sought to maintain a minimum bin population of N = 20.  The second
column lists the mean [Fe/H] of the stars in the listed abundance interval,
and the third lists the numbers of stars involved.  The fourth column lists
$S_{BI}$, along with errors obtained by analysis of 1000 bootstrap
re-samples of the data in the bin.  These errors are useful for assessing the
significance of the difference between the scales of the data from bin to
bin.

In Figure~10 we plot the scale estimates for each set of elemental abundance
ratios listed in Table~8, along with error bars, as a function of
[Fe/H].  In addition to these points, we have also shown a higher-resolution
summary line, obtained by determination of $S_{BI}$ as a function of [Fe/H] for
adjoining bins with a fixed number of stars per bin; for the majority of the
plots, N = 15 was employed, though for several of the plots with a lower
density of points, N=10, was used to preserve resolution.  These lines serve
to guide the eye as to the general behavior of the scale at any given [Fe/H]. 

Our results agree well with those of Fulbright (1999, Chapter 6) in the
abundance ranges he considered, and we refer the reader to his discussion of
the dispersions.  In brief summary, concerning elements in common between the
two works, he concluded from the data set he assembled that the scatter in
Mg, Al, Si is probably real, while for Ca, V, Cr, Ni, and Ba the spread was
not larger than might be expected from the observational errors alone.  In
Table~8 one sees almost invariably that the value of $S_{BI}$ is indeed
larger in the lowest abundance bin than in higher abundance ones. The
question then arises: does the abundance spread increase significantly at
lowest abundance?  If one excludes Si (poor data), and Sr and Ba (clear large
abundance spreads) and considers the nine remaining elements in Table~8 in
the range [Fe/H] $<$ --2.5, one finds $\langle$$S_{BI}$$\rangle$ = 0.19,
which may be compared with the mean observational error from Table~5 of 0.14.
Given, however, that the sample we used to determine $S_{BI}$ is derived from
a large compilation of data with various figures of merit and analyzed by
different workers, the latter value is probably an underestimate of the
appropriate observational error.  It would thus be premature to conclude that
for [Fe/H] $<$ --2.5 the observed scatter exceeds the observational errors.
That said, we also note that the scale estimates for stars in the lowest
abundance bins are indeed significantly higher than those in the highest
abundance bins when compared with the estimated bootstrap errors on the
scale.  We can offer no resolution of this impasse.  It should be resolved
with a much larger and homogeneous data set, both to improve the quality of
the data and to understand the role of unusual stars such as CS~22949--037.

\subsection{Relative Abundance as a Function of Atomic Number}

It is clear from the previous discussion that the abundance trends reported
here are not all well-predicted by canonical Galactic chemical enrichment
models, and that in several cases the input supernova models are inadequate
for the task.  Unexplained trends include the behavior of C, N, Cr, Mn, Co,
and the heavy neutron-capture elements, together with the apparent
Fe-peak-element deficiency of CS~22949--037.  To obtain further insight into
the problem we consider relative abundances as a function of atomic number.

\subsubsection{CS~22949--037}

The upper panel of Figure~11 presents relative abundances for CS~22949--037
relative to the averages of the other four stars studied here, as a function
of atomic number.  As discussed above, the light elements C, N, Mg, and Si
are strongly enhanced relative to those of the iron peak in this
object. (Nitrogen does not appear in Figure~11 since it was not detected in
the other stars of the sample.)

Insofar as C is enhanced in CS~22885--096, one also might wonder if its
abundance pattern is similar to that of CS~22949--037, albeit in milder
degree.  The lower panel of Figure~11 shows the abundances of CS~22885--096
relative to the average of CD--24$^{\rm o}$17504, CD--38$^{\rm o}$245, and
CS~22172--002.  One sees that C, Mg, Al, Si, and Ca all lie above zero as in
CS~22949--037, but the comparison is barely suggestive of the existence of
the phenomenon and hardly compelling, given measurement errors of order 0.15
dex.

An ordinary runs test (see, e.g., Gibbons 1985) can be used to attach
estimates of the statistical significance to the impressions described above.
In the ordered sequence of increasing atomic number, the set of abundances
for CS~22939--037 are all positive with respect to the average of the other
four stars, until one reaches Sc, after which they all remain negative.  The
statistical likelihood of there occurring only two ``runs'' of abundances in a
sample of 12 different species is quite small.  The formal (one-sided)
p-value returned by the runs test is 0.0015, highly significant, and strongly
suggests that the lighter elements are indeed clustered in a positive sense,
and that the result is not arising from chance.  The same test applied to the
case of CS~22885--096, where five separate runs are seen among a sample of 12
species, returns a p-value which is not statistically significant.

\subsubsection{Comparison with Supernovae Models (M $\lesssim$ 100 M$_{\odot}$)} 

Several authors have advocated that the range in relative abundances seen at
lowest abundance is the signature of chemical pre-enrichment by the ejecta of
individual supernovae rather than those from a complete population,
well-mixed into the interstellar medium.  In the present work, CS~22949--037
is a good example of a star which invites such interpretation.
CS~22892--052, with its enormous r-process enhancement (Sneden et al. 1996),
is another.  If the suggestion is valid, it offers the hope of providing
considerable insight into the ejecta of individual supernovae and strong
constraints on models of the phenomenon.  An example of the approach was
given for CS~22876--032 ([Fe/H] = --3.72) by Norris et al. (2000; see their
Figure~8), who argued that the relative proportions of the $\alpha$ elements
Mg, Si, and Ca in this star were consistent with its having been formed from
material which had been preferentially enriched by the ejecta of a 30
M$_{\odot}$, zero-heavy element progenitor.

We sought to address this question in terms of the low abundance models of
Woosley \& Weaver (1995).  Given that the predicted production of the
iron-peak elements is critically sensitive to the assumed ejection energy and
mass cut of the model, together with the overdeficiency of Fe in
CS~22949--037, we chose to normalize relative abundances to magnesium for
both observation and theory.  We were particularly interested to determine
the sensitivity of model predictions to initial abundances, stellar mass, and
assumed ejection energy.  We summarize our conclusions as follows. (1) There
is essentially no difference in the relative abundances of the predictions
for high-mass models having heavy element abundances Z = 0 and Z = 10$^{-4}$,
which simplifies interpretation in terms of Population II or putative
Population III (Z = 0, by definition here) objects. (2) In the range 30--40
M$_{\odot}$, for models having ejection energies sufficiently high that iron
was ejected, the predicted relative abundances are essentially identical.  We
conclude then that insofar as enrichment is produced by stars in this mass
range one might not expect to be able to discriminate progenitor masses to
better than 10 M$_{\odot}$. (3) The models are extremely sensitive to assumed
ejection energy.

Figure~12 compares the element mass fractions in CS~22949--037 relative to
that of Mg with the mean value for the other four stars observed here, as a
function of atomic species.  The predictions of models Z35B and Z35C of
Woosley \& Weaver (1995) are also shown.  (The ordinate is the logarithmic
mass fraction relative to that of Mg.)  Model Z35C has sufficient energy to
eject significant amounts of Fe, while Z35B does not.  The important point of
this figure is that the difference between the data for CS~22949--037 and the
average of the four ``normal" stars is considerably smaller than that between
Z35B and Z35C.  It seems therefore incorrect to suggest that the abundance
patterns of CS~22949--037 are representative of the undiluted ejecta of a
supernova {\it like Z35B}. Given the great sensitivity to energy, it would be
interesting to know whether any models with explosive energies intermediate
to cases B and C reproduce the observations.

If one contrives, however, to represent the abundance patterns of
CS~22949--037 in terms of the hypothesis that it formed from an admixture of
the ejecta of a Z35B-like object with an interstellar medium similar to that
from which the other stars of the sample formed, the extra assumption leads
to considerably better agreement. Table~9 compares the production factors of
Z35B of Woosley and Weaver (1995, their Table~17B) with those of the excess
mass fractions of the elements in CS~22949--037 above the corresponding
average values in the other four stars\footnote{For the four star average we
adjust all [X/H] by 0.16 dex, which brings its value of [Fe/H] into line with
that of CS~22949--037.}, normalized to the Mg production factor (3.34) of
Z35B.  The agreement is far from perfect, with Si through Ti being
poorly-explained.  Given, however, the approximate nature of the modeling the
hypotheses is not preposterous, and deserves closer consideration.

\subsubsection{Comparison with Hypernovae (M $\gtrsim$ 100 M$_{\odot}$
Pair-Instability Supernovae)}

One of the unexplained features of the present results, and indeed of studies
of extremely metal-poor stars in general, is the high incidence of supersolar
[C/Fe] values.  Add to this the supersolar value of [N/Fe] in CS~22949--037.
These features are not predicted by canonical supernovae and GCE models
(Timmes et al. 1995), which predict that [C/Fe] will have roughly the solar
value and that [N/Fe] should be subsolar (since N is produced in a secondary
manner).  While one can suggest {\it ad hoc} explanations yielding supersolar
N involving rotation (e.g. Maeder 1997) and convective overshoot or
supermixing (e.g. Timmes et al. 1995), there exists a class of models which
actually predicted the possibility of supersolar C and N abundances, together
with under-production of the iron-peak group.  These are the hypernovae
(massive pair-instability supernovae) discussed by Woosley \& Weaver (1982).
While these authors emphasize the simplifying assumptions in their models
(for example the neglect of possible mass loss) the results for their 500
M$_{\odot}$, zero-heavy-element hypernova is of particular interest.  This
object evolves to produce large amounts of primary nitrogen by proton capture
on dredged-up carbon, while no Fe will be produced since ``The carbon-oxygen
core itself will certainly become a black hole."  As noted by Carr, Bond, \&
Arnett (1984), the essential feature of these very massive objects is their
potential to ``pass carbon and oxygen from the helium-burning core through
the hydrogen-burning shell, in such a way that it is CNO processed to
nitrogen before entering the hydrogen envelope."  The more recent
computations of Heger \& Woosley (2000, reported by Fryer, Woosley, \& Heger
2000) find large primary nitrogen production in rotating 250 and 300
M$_{\odot}$, Z = 0 hypernovae.  Magnesium enhancement is also reported.

Although the oxygen abundance of CS~22949--037 is unknown, it is likely that
this object contains more nitrogen than all other metals combined.  Given the
unexplained abundance results discussed here (not to mention the existence of
the very poorly-understood object CS~22957--027 noted above ([Fe/H] =
--3.38, [C/Fe] = +2.2, [N/Fe] = 2.0, but ``normal'' heavy neutron-capture
element abundance; see Norris et al. 1997b), systematic theoretical
investigations of hypernovae and their incorporation into models of early
Galactic chemical enrichment might well prove fruitful.

\section{CONCLUSIONS}

We have obtained high-resolution (R = 42000), high-signal-to-noise
($\langle$S/N$\rangle$ = 85) spectra for five stars with [Fe/H] $\lesssim$
--3.5, but otherwise chosen by selection criteria blind to relative abundance
peculiarities.  The data were analyzed with LTE 1D model-atmosphere
techniques to determine chemical abundances (or limits) for some 18 elements.
For atomic species, the accuracy of the line-strength measurements, $\sim$
5--10 m{\AA}, the uncertainties in atmospheric parameters, and the relatively
small numbers of lines available combine to yield a median relative abundance
error of 0.15 dex.  The mean abundance of the group is
$\langle$[Fe/H]$\rangle$ = --3.68 (with $\sigma_{\rm [Fe/H]}$ = 0.23), which
permits one to investigate abundance trends and spreads at lowest abundance.

CD--38$^{\rm o}$245, discovered some two decades ago (Bessell \& Norris
1984), with a newly determined metallicity of [Fe/H] = --3.98, remains the
most metal-deficient object currently known despite comprehensive searches
throughout the intervening period.

For most of the elements, the relative abundances of the sample show little
spread, and hence yield lowest abundance anchor points for relative abundance
trends.  The remarkable exception to this general behavior is CS~22949--037,
which has [Fe/H] = --3.79.  We confirm the result of McWilliam et al. (1995b)
that [C/Fe], [Mg/Fe], and [Si/Fe] all have values $\sim$~1.0, and stand well
above those observed in other stars at this metallicity.  In this star, we
also find [N/Fe] = 2.7.  For CS~22949--037, the abundance patterns are
suggestive of enrichment scenarios involving partial ejection of supernova
mantles (Woosley \& Weaver 1995) and/or massive (200--500 M${_\odot}$)
zero-heavy-element hypernovae (e.g. Woosley \& Weaver 1982).

The other exceptions are the behavior of carbon (Figure 4) and strontium (but
not barium) (Figure 9), which show large spreads at lowest abundances.  The
large range in carbon for [Fe/H] $<$ --3.0 is suggestive of the need for
non-canonical enrichment sources (e.g. hypernovae).  An explanation of the
contrast in the [Sr/Fe] and [Ba/Fe] spreads may require the operation of more
than one process for the production of the heavy neutron-capture elements in
massive stars (Blake et al. 2001; Ryan et al. 2001).

If one removes CS~22949--037 from the discussion, there is no strong upward
trend in [$\alpha$/Fe] as a function of [Fe/H] (Figure 5), in
contradistinction to some reports of the behavior of [O/Fe].  It remains to
be seen whether models of Galactic chemical enrichment can be produced which
will reconcile such different behavior.  To our knowledge, none exist at the
present time.

The analysis of aluminum lines requires inclusion of non-LTE effects
(Baum\"{u}ller \& Gehren 1997).  When approximate correction is made to the
present LTE results we find that the behavior of [Al/Mg] versus [Fe/H] is in
reasonable accord with the GCE models of Timmes et al. (1995).

We confirm the strong and unexpected behavior of the iron-peak abundance
trends of [Cr/Fe], [Mn/Fe], and [Co/Fe] for [Fe/H] $<$ --3.0 reported by
McWilliam et al. (1995b) and Ryan et al. (1996).  In contrast, there is no
trend for [Ni/Fe]: at [Fe/H] $\sim$ --3.7, [Ni/Fe] = +0.08 $\pm$ 0.06.  This
appears to be inconsistent with supernova models of Nakamura et al. (1999)
that explain the observed behavior of the other Fe-peak elements by varying
the position of the model mass cut relative to the Si-burning regions.

Finally, there is suggestive evidence that the spread in abundance increases
with decreasing abundance for essentially all of the elements considered in
the present study.  The reality of this effect requires re-investigation with
high-quality data of a larger sample of objects than is currently available.

\acknowledgements

We are grateful to the Director and staff of the Anglo-Australian
Observatory, and the Australian Time Allocation Committee for providing the
observational facilities used in this study.  We wish to record special
thanks for the persistence of numerous ING staff who, over four observing
sessions, obtained the WHT spectrum of CS~22949--037.  T.C.B. acknowledges
support of NSF grants AST 92--22326, AST 95--29454, and INT 94--17547.

\appendix
 
\section{[Co/Fe] literature values} 

We have identified a probable 0.4--0.5~dex error in the oscillator strength
of the \ion{Co}{1} 4118.8~\AA\ line which dominates the Gratton \& Sneden (1990,
1991) Co analysis of metal-poor stars. We compared the log~$gf$ values from
three sources: laboratory values from Nitz et al. (1999) and Cardon et
al. (1982), and solar values from Gratton \& Sneden (1990). For lines used in
recent stellar analyses, laboratory values show only a small mean difference
$\langle$CSSTW82 $-$ NKWL99$\rangle = -0.023$~dex from four lines, with a
sample standard deviation $\sigma$ = 0.029. (Nitz et al. compare a larger
line set which also shows excellent agreement.) With the exception of the
Co~4118.8~\AA\ line, the differences between the Gratton \& Sneden and Cardon
et al. values are also small: the mean difference for the other five lines in
common is $\langle$GS90$-$CSSTW82$\rangle$ = +0.064 ($\sigma$ = 0.083), but
the difference for the 4118.8~\AA\ line is +0.47~dex. Nitz et al. and Cardon
et al.  measure this line with small formal errors and find log~$gf = -0.47
\pm 0.02$ and $-0.49 \pm 0.08$ respectively, whereas Gratton \& Sneden (1990)
list it as $-$0.02. On the basis of the small formal errors and the
consistency of the two laboratory studies, we conclude that Gratton \&
Sneden's value is too large by a factor of $\simeq$3.

There are two possible remedies: (1) correct the 4118.8~\AA\ log~$gf$ value
to the laboratory scale of, say, Cardon et al.  by adding $-0.47$ to the
Gratton \& Sneden value, or (2) correct the 4118.8~\AA\ log~$gf$ value to the
same solar scale as GS91's other lines, which apparently differs from the
Cardon et al. scale by 0.064~dex.  We elect to do the latter, to keep the
Gratton \& Sneden measurements on their adopted solar scale, but corrected
for the substantial error in the $gf$ value of the 4118.8~\AA\ line. We thus
add $-0.47+0.064 = -0.41$ to the Gratton \& Sneden log~$gf$ value.

Gratton \& Sneden measured the 4118.8~\AA\ line in only five stars. In
HD~122563 and HD~140283 it was the only Co line, whereas in the other stars
there were additional Co lines: HD~64606 has eight in total, HD~134169 has
four, and HD~165195 has three. The corrections we apply to their published Co
abundances are: HD~64606 +0.05~dex, HD~122563 +0.41, HD~134169 +0.10,
HD~140283 +0.41, and HD~165195 +0.14. We note that the impact of the error
diminishes in more metal-rich stars, due to averaging over more lines.  That
is, the effect of the error is metallicity dependent.

\clearpage
%TABLE1
\begin{center}
\begin{deluxetable}{lccrcr}
\tablecaption{RECENT INVESTIGATIONS OF STARS HAVING [Fe/H] $<$ --3.0}
\tablehead{
\colhead{Investigator}& {R}& {$\langle$S/N$\rangle$}& {N\tablenotemark{a}} & {$\lambda$} & {F}\\
         {}&             {} & {(per pixel)}          &     & {(\AA)}     & {(\AA)$^{-1}$}
\\ 
        {(1)} &{(2)} &{(3)} & {(4)} &{(5)} & {(6) } 
}
\startdata
Molaro \& Bonifacio 1990        &    20000 &  20  & 2  &   4700  &   85 \\
Molaro \& Castelli 1990         &    15000 &  75  & 2  &   4700  &  239 \\    
Ryan, Norris, \& Bessell 1991   &    54000 &  40  & 8  &   4300  &  502 \\
Norris, Peterson, \& Beers 1993 &    50000 &  25  & 5  &   4300  &  290 \\
Primas et al. 1994              &    26000 &  45  & 5  &   4500  &  260 \\
McWilliam et al. 1995a          &    22000 &  35  & 14 &   4800  &  160 \\
Norris, Ryan, \& Beers 1996     &    40000 &  45  & 10 &   4300  &  419 \\\\
This work                       &    42000 &  85  & 5  &   4300  &  830 \\
\enddata
\tablenotetext{a} {Number of stars having [Fe/H] $<$ --3.0.}
\end{deluxetable}
\end{center}

%TABLE2
\begin{center}
\tabletypesize{\scriptsize}
\begin{deluxetable}{crrrrrrr}
\tablecaption{EQUIVALENT WIDTHS FOR PROGRAM STARS}
\tablehead{
\multicolumn{3}{c}{}&\multicolumn{1}{c}{CD--24$^{\rm o}$17504}&\multicolumn{1}{c}{CD--38$^{\rm o}$245}&\multicolumn{1}{c}{CS~22172--002}&\multicolumn{1}{c}{CS~22885--096}&\multicolumn{1}{c}{CS~22949--037}\\
\colhead{$\lambda$}    & {$\chi$}    &{log~$gf$\tablenotemark{a}}& {W$_{\lambda}$}& {W$_{\lambda}$}& {W$_{\lambda}$}& {W$_{\lambda}$} & {W$_{\lambda}$}\\ 
 {(\AA)} & {($\rm eV$)}&       & {(m\AA)}  & {(m\AA)}  & {(m\AA)} & {(m\AA)}& {(m\AA)}  \\
  {(1)}  &    {(2)}    & {(3)} &  {(4)}    &  {(5)}    &  {(6)}   &  {(7)}  &  {(8)}
}
\startdata
Mg I    &         &          &         &         &         &        \\
3829.35 &   2.71  & $-$0.48  &     75  &     95  &    106  &    103  &     146\\
3832.30 &   2.71  & $-$0.13  &    ...  &    111  &    123  &    117  &     164\\
3838.30 &   2.72  & $-$0.10  &    ...  &    108  &    127  &    129  &     162\\
4057.51 &   4.35  & $-$0.89  &   $<$4  &   $<$5  &   $<$8  &   $<$5  &     ...\\
4351.91 &   4.35  & $-$0.56  &     15  &     13  &     24  &     17  &      53\\
4571.10 &   0.00  & $-$5.61  &   $<$4  &   $<$5  &   $<$8  &   $<$5  &      30\\
4703.00 &   4.35  & $-$0.38  &      8  &      7  &     13  &     14  &      46\\
Al I    &         &          &         &         &         &         &  \\
3944.01 &   0.00  & $-$0.64  &     21  &     45  &     70  &     65  &     111\\
3961.52 &   0.01  & $-$0.34  &     25  &     56  &     66  &     62  &      81\\
Si I    &         &          &         &         &         &        \\
3905.52 &   1.91  & $-$1.09  &     57  &     92  &    115  &    116  &     156\\
4102.94\tablenotemark{b} &   1.91  & $-$3.10  &     &     &     &     &        \\
Ca I    &         &          &         &         &         &        \\
4226.73 &   0.00  &   +0.24  &     75  &    103  &    116  &    113  &     119\\
4283.01 &   1.89  & $-$0.22  &   $<$4  &     10  &   $<$8  &      9  &      25\\
4289.36 &   1.88  & $-$0.30  &   $<$4  &   $<$5  &   $<$8  &      9  &     ...\\
4302.53 &   1.90  &   +0.28  &      9  &      9  &     32  &    ...  &     ...\\
4318.65 &   1.90  & $-$0.21  &      6  &      5  &      9  &     10  &     ...\\
4434.96 &   1.89  & $-$0.01  &      6  &      8  &     16  &     15  &     ...\\
4454.78 &   1.90  &   +0.26  &     10  &      9  &     24  &     20  &      18\\
4455.89 &   1.90  & $-$0.53  &   $<$4  &   $<$5  &   $<$8  &    ...  &   $<$10\\
Sc II   &         &          &         &         &         &        \\
4246.82 &   0.32  &   +0.24  &     20  &     55  &     66  &     67  &      57\\
4294.78 &   0.61  & $-$1.39  &   $<$4  &   $<$5  &   $<$8  &   $<$5  &     ...\\
4314.08 &   0.62  & $-$0.10  &      6  &     19  &     35  &     30  &      43\\
4320.73 &   0.61  & $-$0.25  &      7  &     15  &     25  &     22  &     ...\\
4324.99 &   0.60  & $-$0.44  &   $<$4  &     10  &     18  &     17  &   $<$10\\
4400.39 &   0.61  & $-$0.54  &   $<$4  &      9  &     10  &     13  &      13\\
4415.55 &   0.60  & $-$0.67  &   $<$4  &      9  &     14  &     15  &      12\\
Ti I    &         &          &         &         &         &        \\
3924.53 &   0.02  & $-$0.88  &   $<$4  &   $<$5  &   $<$8  &   $<$5  &   $<$10\\
3958.21 &   0.05  & $-$0.12  &   $<$4  &     13  &    ...  &     10  &     ...\\
3989.76 &   0.02  & $-$0.14  &   $<$4  &    ...  &     31  &     13  &     ...\\
3998.64 &   0.05  &   +0.00  &   $<$4  &     11  &     21  &     13  &   $<$10\\
4533.24 &   0.85  &   +0.53  &   $<$4  &   $<$5  &      8  &      7  &   $<$10\\
Ti II   &         &          &         &         &         &        \\
3741.63 &   1.58  & $-$0.11  &     17  &     39  &    ...  &     37  &     ...\\
3757.68 &   1.57  & $-$0.46  &   $<$8  &     25  &    ...  &     24  &     ...\\
3759.30 &   0.61  &   +0.27  &     70  &    121  &    110  &    109  &     134\\
3761.32 &   0.57  &   +0.17  &     69  &    114  &    113  &    116  &     117\\
3813.34 &   0.61  & $-$2.02  &   $<$8  &     22  &     34  &     22  &     ...\\
3900.55 &   1.13  & $-$0.45  &     27  &     71  &     85  &     62  &      72\\
3913.47 &   1.12  & $-$0.53  &     24  &     62  &     73  &     61  &      72\\
3987.63 &   0.61  & $-$2.73  &   $<$4  &   $<$5  &   $<$8  &   $<$5  &   $<$10\\
4012.39 &   0.57  & $-$1.75  &   $<$4  &     29  &     48  &     27  &      35\\
4025.13 &   0.61  & $-$1.98  &   $<$4  &     12  &     26  &     10  &      12\\
4028.33 &   1.89  & $-$1.00  &   $<$4  &   $<$5  &   $<$8  &      5  &     ...\\
4173.54 &   1.08  & $-$2.00  &   $<$4  &      8  &     13  &   $<$5  &   $<$10\\
4287.87 &   1.08  & $-$2.02  &   $<$4  &      7  &     10  &      5  &   $<$10\\
4290.22 &   1.17  & $-$1.12  &      8  &     27  &     45  &     26  &      30\\
4300.05 &   1.18  & $-$0.49  &     16  &     42  &     59  &     39  &     ...\\
4301.94 &   1.16  & $-$1.20  &   $<$4  &     19  &     35  &     23  &     ...\\
4312.86 &   1.18  & $-$1.16  &      5  &     18  &     46  &    ...  &     ...\\
4330.24 &   2.05  & $-$1.51  &   $<$4  &   $<$5  &   $<$8  &   $<$5  &   $<$10\\
4330.70 &   1.18  & $-$2.06  &   $<$4  &   $<$5  &   $<$8  &   $<$5  &   $<$10\\
4337.92 &   1.08  & $-$1.13  &   $<$4  &     33  &     59  &    ...  &      58\\
4394.06 &   1.22  & $-$1.77  &   $<$4  &   $<$5  &   $<$8  &   $<$5  &   $<$10\\
4395.03 &   1.08  & $-$0.51  &     19  &     55  &     67  &     52  &      60\\
4399.77 &   1.24  & $-$1.27  &      5  &     13  &     26  &     15  &      20\\
4417.72 &   1.17  & $-$1.43  &   $<$4  &     20  &     34  &     17  &      24\\
4418.34 &   1.24  & $-$1.99  &   $<$4  &   $<$5  &   $<$8  &   $<$5  &   $<$10\\
4443.80 &   1.08  & $-$0.70  &     16  &     47  &     64  &     44  &      56\\
4444.55 &   1.12  & $-$2.21  &   $<$4  &   $<$5  &   $<$8  &   $<$5  &   $<$10\\
4450.48 &   1.08  & $-$1.51  &      8  &     13  &     28  &     14  &      19\\
4464.46 &   1.16  & $-$2.08  &   $<$4  &      6  &   $<$8  &      8x\tablenotemark{c} &     23x\tablenotemark{c}\\
4468.51 &   1.13  & $-$0.60  &     18  &     45  &     59  &     34  &      65\\
4470.87 &   1.17  & $-$2.28  &   $<$4  &   $<$5  &   $<$8  &      6x\tablenotemark{c} &   $<$10\\
4501.27 &   1.12  & $-$0.76  &     13  &     46  &     57  &     39  &      47\\
4533.96 &   1.24  & $-$0.77  &     14  &     42  &     59  &     40  &      44\\
4563.76 &   1.22  & $-$0.96  &      7  &     31  &     47  &     27  &      40\\
4571.97 &   1.57  & $-$0.53  &     13  &     31  &     43  &     28  &      48\\
4589.96 &   1.24  & $-$1.79  &   $<$4  &   $<$5  &   $<$8  &      7  &     ...\\
V II    &         &          &         &         &         &        \\
3951.96 &   1.48  & $-$0.78  &   $<$4  &   $<$5  &    ...  &   $<$5  &   $<$10\\
Cr I    &         &          &         &         &         &        \\
3991.12 &   2.55  &   +0.25  &   $<$4  &   $<$5  &    ...  &   $<$5  &   $<$10\\
4254.33 &   0.00  & $-$0.11  &     21  &     35  &     48  &     38  &      43\\
4274.80 &   0.00  & $-$0.23  &     19  &     32  &     56  &     37  &      38\\
4289.72 &   0.00  & $-$0.36  &     13  &     23  &     39  &     31  &      40\\
Mn I    &         &          &         &         &         &        \\
4030.75 &   0.00  & $-$0.47  &     16  &     17  &     34  &     34  &      25\\
4033.06 &   0.00  & $-$0.62  &     10  &     14  &     21  &     25  &      32\\
4034.48 &   0.00  & $-$0.81  &      8  &     14  &     23  &     25  &     ...\\
Fe I    &         &          &         &         &         &        \\
3727.63 &   0.96  & $-$0.62  &     50  &     91  &    ...  &     98  &     ...\\
3743.37 &   0.99  & $-$0.78  &     47  &     81  &    ...  &     92  &     ...\\
3745.57 &   0.09  & $-$0.77  &    ...  &    141  &    ...  &    135  &     ...\\
3745.91 &   0.12  & $-$1.34  &    ...  &    125  &    ...  &    110  &     ...\\
3748.27 &   0.11  & $-$1.01  &    ...  &    ...  &    ...  &    ...  &     ...\\
3758.24 &   0.96  & $-$0.02  &     73  &    127  &    ...  &    110  &      92\\
3763.80 &   0.99  & $-$0.23  &    ...  &    102  &    120  &     95  &     103\\
3765.54 &   3.24  &   +0.48  &    ...  &    ...  &     27  &     18  &     ...\\
3767.20 &   1.01  & $-$0.39  &    ...  &     92  &    109  &     98  &      94\\
3786.67 &   1.01  & $-$2.19  &    ...  &     16  &    ...  &     22  &     ...\\
3787.88 &   1.01  & $-$0.85  &     45  &     84  &    106  &     88  &      79\\
3790.09 &   0.99  & $-$1.74  &    ...  &     42  &     75  &     48  &     ...\\
3795.00 &   0.99  & $-$0.75  &    ...  &     98  &    101  &     85  &     112\\
3805.34 &   3.30  &   +0.31  &      9  &     12  &     19  &     15  &   $<$15\\
3807.54 &   2.22  & $-$0.99  &      8  &     14  &    ...  &     18  &   $<$15\\
3808.73 &   2.56  & $-$1.14  &   $<$8  &  $<$10  &    ...  &  $<$10  &   $<$15\\
3812.96 &   0.96  & $-$1.03  &     38  &     79  &     91  &     82  &      90\\
3815.84 &   1.49  &   +0.24  &     69  &     98  &    113  &     95  &     101\\
3820.43 &   0.86  &   +0.14  &     84  &    130  &    135  &    124  &     125\\
3821.19 &   3.27  &   +0.20  &      9  &     11  &     32  &     12  &   $<$15\\
3824.44 &   0.00  & $-$1.35  &     67  &    129  &    139  &    107  &     131\\
3825.88 &   0.92  & $-$0.03  &     77  &    117  &    133  &    116  &     126\\
3827.82 &   1.56  &   +0.08  &     61  &     90  &    106  &     96  &      85\\
3839.26 &   3.05  & $-$0.33  &    ...  &  $<$10  &     17  &  $<$10  &     ...\\
3840.44 &   0.99  & $-$0.50  &    ...  &     89  &    105  &     92  &      84\\
3849.97 &   1.01  & $-$0.87  &     40  &     79  &     93  &     73  &     120\\
3850.82 &   0.99  & $-$1.74  &     13  &     45  &     58  &     46  &     ...\\
3852.58 &   2.18  & $-$1.19  &   $<$8  &  $<$10  &    ...  &  $<$10  &     ...\\
3856.37 &   0.05  & $-$1.28  &     69  &    132  &    129  &    118  &      88\\
3859.21 &   2.41  & $-$0.75  &   $<$8  &     13  &     31  &     22  &     ...\\
3859.91 &   0.00  & $-$0.70  &     89  &    146  &    148  &    143  &     130\\
3865.52 &   1.01  & $-$0.97  &     36  &     75  &     82  &     85  &      67\\
3867.22 &   3.02  & $-$0.45  &   $<$8  &  $<$10  &    ...  &  $<$10  &     ...\\
3871.75 &   2.95  & $-$0.84  &   $<$8  &  $<$10  &    ...  &    ...  &     ...\\
3872.50 &   0.99  & $-$0.91  &     40  &     89  &     97  &     81  &      72\\
3876.04 &   1.01  & $-$2.86  &   $<$8  &  $<$10  &  $<$15  &  $<$10  &   $<$15\\
3878.02 &   0.96  & $-$0.91  &     39  &     85  &     98  &     90  &      89\\
3878.57 &   0.09  & $-$1.36  &     65  &    126  &    139  &    119  &     133\\
3885.51 &   2.43  & $-$1.09  &   $<$8  &  $<$10  &  $<$15  &  $<$10  &   $<$15\\
3886.28 &   0.05  & $-$1.07  &    ...  &    124  &    131  &    125  &     129\\
3887.05 &   0.92  & $-$1.12  &    ...  &     83  &    100  &     85  &      80\\
3895.66 &   0.11  & $-$1.66  &    ...  &     97  &    112  &    107  &     105\\
3898.01 &   1.01  & $-$2.02  &    ...  &     34  &     52  &     46  &     ...\\
3899.71 &   0.09  & $-$1.52  &     57  &    117  &    134  &    107  &     108\\
3902.95 &   1.56  & $-$0.44  &     37  &     68  &     80  &     65  &      63\\
3906.48 &   0.11  & $-$2.20  &     20  &     85  &     86  &     81  &      71\\
3916.72 &   3.24  & $-$0.58  &   $<$4  &   $<$5  &   $<$8  &   $<$5  &   $<$10\\
3917.18 &   0.99  & $-$2.15  &      6  &     25  &     49  &     32  &      29\\
3920.26 &   0.12  & $-$1.74  &     43  &    106  &    108  &     91  &     112\\
3922.91 &   0.05  & $-$1.64  &     54  &    114  &    119  &    103  &     115\\
3927.92 &   0.11  & $-$1.52  &     53  &    115  &    125  &    109  &     139\\
3930.30 &   0.09  & $-$1.49  &     60  &    112  &    ...  &     99  &     121\\
3940.88 &   0.96  & $-$2.55  &   $<$4  &      9  &     25  &     17  &   $<$10\\
3949.95 &   2.18  & $-$1.25  &   $<$4  &      6  &     23  &      9  &     ...\\
3983.96 &   2.73  & $-$1.02  &   $<$4  &   $<$5  &   $<$8  &   $<$5  &   $<$10\\
3997.39 &   2.73  & $-$0.48  &   $<$4  &      8  &    ...  &     18  &   $<$10\\
4005.24 &   1.56  & $-$0.60  &     31  &     68  &     88  &     67  &      66\\
4009.72 &   2.22  & $-$1.25  &   $<$4  &      8  &     12  &     10  &   $<$10\\
4045.81 &   1.49  &   +0.28  &     72  &    102  &    123  &    108  &     108\\
4062.44 &   2.85  & $-$0.86  &   $<$4  &   $<$5  &   $<$8  &   $<$5  &   $<$10\\
4063.59 &   1.56  &   +0.06  &     60  &     92  &     97  &     93  &      90\\
4067.97 &   3.21  & $-$0.47  &   $<$4  &   $<$5  &   $<$8  &   $<$5  &   $<$10\\
4071.74 &   1.61  & $-$0.02  &     54  &     83  &     97  &     81  &      86\\
4076.62 &   3.21  & $-$0.53  &   $<$4  &   $<$5  &    ...  &   $<$5  &   $<$10\\
4084.49 &   3.33  & $-$0.71  &   $<$4  &   $<$5  &   $<$8  &   $<$5  &   $<$10\\
4132.06 &   1.61  & $-$0.68  &     28  &     60  &     75  &     64  &      66\\
4132.90 &   2.85  & $-$1.01  &   $<$4  &   $<$5  &   $<$8  &   $<$5  &   $<$10\\
4134.68 &   2.83  & $-$0.65  &   $<$4  &   $<$5  &     18  &   $<$5  &   $<$10\\
4137.00 &   3.42  & $-$0.45  &   $<$4  &   $<$5  &   $<$8  &   $<$5  &   $<$10\\
4143.42 &   3.05  & $-$0.20  &   $<$4  &      8  &     17  &     13  &   $<$10\\
4143.87 &   1.56  & $-$0.51  &     34  &     70  &     88  &     74  &      75\\
4147.67 &   1.49  & $-$2.09  &   $<$4  &      6  &     14  &     11  &   $<$10\\
4154.50 &   2.83  & $-$0.69  &   $<$4  &   $<$5  &   $<$8  &   $<$5  &   $<$10\\
4156.80 &   2.83  & $-$0.81  &   $<$4  &   $<$5  &   $<$8  &   $<$5  &   $<$10\\
4157.77 &   3.42  & $-$0.40  &   $<$4  &   $<$5  &   $<$8  &   $<$5  &   $<$10\\
4174.91 &   0.92  & $-$2.95  &   $<$4  &      7  &     17  &      9  &   $<$10\\
4175.64 &   2.85  & $-$0.83  &   $<$4  &   $<$5  &   $<$8  &   $<$5  &   $<$10\\
4181.75 &   2.83  & $-$0.37  &      6  &     11  &     15  &     11  &   $<$10\\
4184.89 &   2.83  & $-$0.87  &   $<$4  &   $<$5  &   $<$8  &   $<$5  &   $<$10\\
4187.04 &   2.45  & $-$0.53  &      8  &     15  &     31  &     22  &      20\\
4187.79 &   2.43  & $-$0.53  &      9  &     16  &     32  &     18  &      23\\
4198.31 &   2.40  & $-$0.67  &      8  &     15  &     30  &     22  &      20\\
4199.10 &   3.05  &   +0.16  &     12  &     18  &     27  &     21  &      27\\
4202.03 &   1.49  & $-$0.70  &     31  &     68  &     84  &     72  &      72\\
4206.70 &   0.05  & $-$3.96  &   $<$4  &     10  &    ...  &     19  &     31x\tablenotemark{c}\\
4210.35 &   2.48  & $-$0.93  &   $<$4  &      9  &     23  &     11  &      15\\
4216.18 &   0.00  & $-$3.36  &   $<$4  &     33  &     48  &     33  &      35\\
4219.36 &   3.58  &   +0.00  &   $<$4  &   $<$5  &   $<$8  &   $<$5  &   $<$10\\
4222.21 &   2.45  & $-$0.94  &   $<$4  &      8  &     17  &      9  &     ...\\
4227.43 &   3.33  &   +0.27  &     10  &     13  &     24  &     15  &      21\\
4233.60 &   2.48  & $-$0.59  &      5  &     15  &     31  &     22  &      26\\
4235.94 &   2.43  & $-$0.33  &     12  &     25  &     53  &     34  &     ...\\
4238.02 &   3.42  & $-$0.62  &   $<$4  &   $<$5  &   $<$8  &      7  &     ...\\
4238.80 &   3.40  & $-$0.23  &   $<$4  &   $<$5  &   $<$8  &      6  &   $<$10\\
4247.42 &   3.37  & $-$0.24  &   $<$4  &   $<$5  &     10  &    ...  &   $<$10\\
4250.12 &   2.47  & $-$0.39  &     11  &     20  &     41  &     27  &      25\\
4250.79 &   1.56  & $-$0.71  &     28  &     60  &     79  &     61  &      56\\
4260.47 &   2.40  &   +0.11  &     26  &     46  &     62  &     48  &      48\\
4266.97 &   2.73  & $-$1.81  &   $<$4  &   $<$5  &   $<$8  &   $<$5  &   $<$10\\
4271.15 &   2.45  & $-$0.34  &     12  &     23  &     39  &     29  &      38\\
4271.76 &   1.49  & $-$0.17  &     56  &     89  &    105  &     76  &      94\\
4282.41 &   2.18  & $-$0.78  &      9  &     17  &     28  &     22  &      22\\
4294.12 &   1.49  & $-$1.04  &     23  &     62  &     82  &     65  &      69\\
4299.23 &   2.43  & $-$0.38  &     12  &     25  &     56  &     40  &     ...\\
4307.90 &   1.56  & $-$0.07  &     60  &     92  &    117  &     94  &     ...\\
4325.76 &   1.61  &   +0.01  &     58  &     83  &     97  &     85  &      86\\
4337.05 &   1.56  & $-$1.70  &    ...  &   $<$5  &     40  &    ...  &     ...\\
4375.93 &   0.00  & $-$3.02  &      5  &     54  &     76  &     51  &      39\\
4383.54 &   1.49  &   +0.20  &     72  &     85x\tablenotemark{c} &    116  &     98  &      98\\
4389.25 &   0.05  & $-$4.57  &   $<$4  &   $<$5  &   $<$8  &   $<$5  &   $<$10\\
4404.75 &   1.56  & $-$0.13  &     57  &     85  &     99  &     89  &      82\\
4415.12 &   1.61  & $-$0.62  &     34  &     56  &     81  &     68  &      72\\
4427.31 &   0.05  & $-$2.92  &      6  &     50  &     74  &     52  &      45\\
4430.61 &   2.22  & $-$1.69  &   $<$4  &   $<$5  &   $<$8  &   $<$5  &   $<$10\\
4442.34 &   2.20  & $-$1.24  &      7  &   $<$5  &     17  &      9  &   $<$10\\
4447.72 &   2.22  & $-$1.34  &   $<$4  &   $<$5  &     11  &      9  &   $<$10\\
4454.39 &   2.83  & $-$1.30  &   $<$4  &   $<$5  &   $<$8  &   $<$5  &   $<$10\\
4459.12 &   2.18  & $-$1.31  &   $<$4  &      9  &     19  &     10  &     ...\\
4461.65 &   0.09  & $-$3.20  &   $<$4  &     39  &     60  &     37  &      38\\
4466.55 &   2.83  & $-$0.60  &   $<$4  &     10  &     14  &      8  &   $<$10\\
4482.17 &   0.11  & $-$3.48  &   $<$4  &     28  &     46  &     24  &      20\\
4489.74 &   0.12  & $-$3.93  &   $<$4  &     10  &     18  &   $<$5  &     ...\\
4494.56 &   2.20  & $-$1.14  &   $<$4  &     10  &     23  &     15  &      19\\
4528.62 &   2.18  & $-$0.85  &      9  &     18  &     29  &     23  &     ...\\
4531.15 &   1.49  & $-$2.13  &   $<$4  &      8  &     19  &     13  &   $<$10\\
4602.94 &   1.49  & $-$2.21  &   $<$4  &      8  &     16  &     13  &   $<$10\\
Fe II   &         &          &         &         &         &        \\
4178.85 &   2.58  & $-$2.48  &   $<$4  &   $<$5  &     10  &      9  &   $<$10\\
4233.16 &   2.58  & $-$1.91  &      8  &     14  &     32  &     24  &     ...\\
4303.17 &   2.71  & $-$2.57  &   $<$4  &   $<$5  &   $<$8  &   $<$5  &     ...\\
4385.37 &   2.78  & $-$2.57  &   $<$4  &   $<$5  &   $<$8  &   $<$5  &     ...\\
4508.28 &   2.86  & $-$2.21  &   $<$4  &   $<$5  &      8  &   $<$5  &   $<$10\\
4515.33 &   2.85  & $-$2.48  &   $<$4  &   $<$5  &   $<$8  &   $<$5  &   $<$10\\
4520.22 &   2.81  & $-$2.60  &   $<$4  &   $<$5  &   $<$8  &   $<$5  &   $<$10\\
4522.62 &   2.85  & $-$2.03  &   $<$4  &   $<$5  &     15  &   $<$5  &      12\\
4555.88 &   2.83  & $-$2.29  &   $<$4  &   $<$5  &      9  &   $<$5  &   $<$10\\
4583.83 &   2.81  & $-$2.02  &      6  &     11  &     26  &     16  &      14\\
Co I    &         &          &         &         &         &        \\
3842.05 &   0.92  & $-$0.77  &    ...  &  $<$10  &     15  &     14  &   $<$15\\
3873.11 &   0.43  & $-$0.66  &     11  &     44  &     68  &     51  &      41\\
3873.96 &   0.51  & $-$0.87  &    ...  &     27  &     52  &     38  &      32\\
3894.07 &   1.05  &   +0.10  &    ...  &     23  &     61  &     35  &   $<$15\\
4121.31 &   0.92  & $-$0.32  &      4  &     21  &     40  &     27  &     ...\\
Ni I    &         &          &         &         &         &        \\
3775.57 &   0.42  & $-$1.41  &     18  &     35  &     74  &     57  &     ...\\
3783.53 &   0.42  & $-$1.31  &     20  &     39  &     65  &     59  &      70\\
3807.14 &   0.42  & $-$1.22  &     22  &     43  &     63  &     60  &      46\\
3831.70 &   0.42  & $-$2.27  &   $<$8  &     14  &     28  &     25  &      21\\
3858.29 &   0.42  & $-$0.95  &     35  &     64  &     74  &     74  &      70\\
Sr II   &         &          &         &         &         &        \\
4077.71 &   0.00  &   +0.15  &   $<$4  &     68  &     58  &     29  &     100\\
4215.52 &   0.00  & $-$0.17  &   $<$4  &     53  &     38  &     16  &      97\\
Y II    &         &          &         &         &         &        \\
3774.33 &   0.12  &   +0.21  &    ...  &  $<$10  &  $<$15  &  $<$10  &     ...\\
3788.70 &   0.10  & $-$0.07  &   $<$8  &  $<$10  &  $<$15  &  $<$10  &      33\\
3818.34 &   0.12  & $-$0.98  &   $<$8  &  $<$10  &  $<$15  &  $<$10  &   $<$15\\
3950.36 &   0.10  & $-$0.49  &   $<$4  &   $<$5  &  $<$15  &   $<$5  &     ...\\
4374.94 &   0.40  &   +0.02  &   $<$4  &   $<$5  &   $<$8  &   $<$5  &      14\\
Ba II   &         &          &         &         &         &        \\
4554.03 &   0.00  &   +0.16  &   $<$4  &     16  &     15  &      6  &      23\\
Eu II   &         &          &         &         &         &        \\
4129.70 &   0.00  &   +0.20  &   $<$4  &   $<$5  &   $<$8  &   $<$5  &   $<$10\\
4205.05 &   0.00  &   +0.12  &   $<$4  &   $<$5  &   $<$8  &   $<$5  &     ...\\
\enddata
\tablenotetext{a} {Sources of $gf$ values may be found in Norris et
al. (1996), except for Sc II, Ti II, and V II which come from Lawler
\& Dakin (1989), (where possible) Bizzarri et al. (1993), and
Karamatskos et al. (1986), respectively.}
\tablenotetext{b}{Spectrum synthesis used for giants.  See text.}
\tablenotetext{c}{Line measured but excluded in abundance analysis due to discrepant abundance.}
\end{deluxetable}
%}
\end{center}

%TABLE3
\clearpage
\begin{center}
\begin{deluxetable}{llrrrr}
\tablecaption{RADIAL VELOCITIES FOR PROGRAM STARS}
\tablehead{
\colhead{Object} & {Date} &{V$_{\rm r}$} &{s.e.\tablenotemark{a}} & {V$_{\rm r}$\tablenotemark{b}}& {V$_{\rm r}\tablenotemark{c}$} \\ 
        {(1)} &{(2)} &{(3)} & {(4)} &{(5)} &{(6)}  }
\startdata
CD--24$^{\rm o}$17504 &1996 Aug 05    &+135.9 &0.03&   ...&  ... \\
CD--38$^{\rm o}$245   &1997 Aug 21    &+45.9  &0.10&+46.9  &+45.7 \\
CS~22172--002         &1998 Aug 14    &+251.3 &0.05&   ... &+250.8\\
CS~22885--096         &1997 Aug 21--22&--251.0 &0.08&--250.1&  ... \\
CS~22949--037         &2000 Sep 5--17 &--125.7&0.19&--126.4&  ... \\
\enddata 
\tablenotetext{a}{Internal error}
\tablenotetext{b}{From McWilliam et al. (1995a)}
\tablenotetext{c}{From Norris et al. (1996)}
\end{deluxetable}
\end{center}

%TABLE4
\clearpage
\begin{center}
\begin{deluxetable}{lccccccccc}
\tablecaption{COLORS AND ATMOSPHERIC PARAMETERS FOR PROGRAM STARS}
\tablehead{
\colhead{Star}& {\it B--V}& {\it V--R}& {\it R--I}&    {E({\it B--V})}& {Source\tablenotemark{a}}& {T$_{\rm eff}$}& {log~$g$}& {[Fe/H]} & {$\xi$}\\
 {}&   {}&   {}&   {}&   {}&   {}&   {}&   {}&   {}&   {(\kms)}\\ 
 {(1)}& {(2)}&  {(3)}&  {(4)}& {(5)}     & {(6)}&    {(7)}&    {(8)}&     {(9)}&     {(10)} 
}
\startdata
CD--24$^{\rm o}$17504&0.39&  ...&0.310&0.00&1&6070 &3.6&--3.37&1.4\\ 
CD--38$^{\rm o}$245  &0.81&  ...&0.505&0.00&2&4850 &1.8&--3.98&2.1\\  
CS~22172--002         &0.81&  ...&0.544&0.06&2&4900 &2.0&--3.61&2.0\\  
CS~22885--096         &0.69&  ...&0.480&0.03&2&5050 &1.9&--3.66&1.9\\ 
CS~22949--037         &0.74& 0.49&  ...&0.03&3&4900 &1.7&--3.79&2.0\\ 
\enddata
\tablenotetext{a} {Sources.---(1) Ryan et al. 1991; (2) Ryan et al. 1996; (3) Preston, Shectman, \& Beers 1991, McWilliam et al. 1995b,  present investigation.}
\end{deluxetable}
\end{center}

%TABLE5
\clearpage
\begin{center}
\tabletypesize{\tiny}
\begin{deluxetable}{llcrrrrrrrrrrrrrrr}
\setlength{\tabcolsep}{0.06in} 
\tablecaption{[Fe/H]\tablenotemark{a} AND RELATIVE ABUNDANCES, [X/Fe]}
\tablehead{
\multicolumn{3}{c}{}&\multicolumn{3}{c}{CD--24$^{\rm o}$17504\tablenotemark{c}}&\multicolumn{3}{c}{CD--38$^{\rm o}$245\tablenotemark{c}}&\multicolumn{3}{c}{CS~22172--002\tablenotemark{c}}&\multicolumn{3}{c}{CS~22885--096\tablenotemark{c}}&\multicolumn{3}{c}{CS~22949--037\tablenotemark{c}}\\
\colhead {Element}& {Species}&  {log~(N/N$_{\rm H}$)$_{\odot}$\tablenotemark{b}}& {[X/Fe]}& {$\sigma$}& {n}& {[X/Fe]}& {$\sigma$}& {n}& {[X/Fe]}& {$\sigma$} & {n}& {[X/Fe]}& {$\sigma$}& {n}&{[X/Fe]}&{$\sigma$}& {n}\\ 
% {(1)}  &{(2)}&    {(3)}&{(4)}&{(5)}   &{(6)}&{(7)}&{(8)}   &{(9)}&{(10)}&{(11)} &{(12)}&{(13)}&{(14)}&{(15)}&{(16)}&{(17)} &{18}
}
\startdata
Fe&Fe I\tablenotemark{a} &  --4.50 &    --3.37& 0.11& 61&    --3.98& 0.15& 92&     --3.61& 0.15& 89&     --3.66& 0.15& 97&--3.79&0.16&61\\
Fe&Fe II&  --4.50 &    --0.02& 0.16&  2&      0.00& 0.25&  2&       0.04& 0.22&  6&     --0.01& 0.23&  3&0.00&0.29&2\\
C & CH& --3.33 &       ...&  ...& ..&   $<$0.00& 0.20& ..&       0.10& 0.20& ..&       0.60& 0.20& ..&1.05&0.20&..\\
N & CN& --4.01 &       ...&  ...& ..&       ...&  ...& ..&        ...&  ...& ..&        ...&  ...& ..&2.70&0.40&..\\
Mg&Mg I &  --4.42 &      0.47& 0.15&  3&      0.45& 0.09&  5&       0.37& 0.09&  5&       0.52& 0.08&  5&1.22&0.14&5\\
Al&Al I &  --5.53 &    --0.79& 0.11&  1\tablenotemark{d}&    --0.77& 0.23&  1\tablenotemark{d}&     --0.94& 0.22&  1\tablenotemark{d}&     --0.78& 0.21&  1\tablenotemark{d}&--0.43&0.32&1\tablenotemark{d}\\
Si&Si I &  --4.45 &      0.04& 0.15&  1&      0.30& 0.21&  2&       0.38& 0.20&  2&       0.44& 0.19&  2&1.04&0.30&2\\
Ca&Ca I &  --5.64 &      0.24& 0.06&  5&      0.14& 0.11&  6&       0.14& 0.10&  5&       0.28& 0.09&  6&0.45&0.21&3\\
Sc&Sc II&  --8.90 &      0.27& 0.16&  3&      0.08& 0.16&  6&       0.03& 0.16&  6&       0.15& 0.16&  6&0.13&0.20&4\\
Ti&Ti I &  --7.01 &   $<$0.44& 0.06&   1\tablenotemark{e}&     0.28& 0.14&  2&       0.24& 0.15&  3&       0.21& 0.09&  4&$<$0.02&0.04&1\\
Ti&Ti II&  --7.01 &      0.27& 0.14& 17&      0.37& 0.15& 27&       0.36& 0.15& 24&       0.12& 0.15& 25&0.35&0.16&18\\
V &V  II&  --8.00 &   $<$0.94& 0.15&  1&   $<$0.52& 0.19&  1&        ...&  ...& ..&    $<$0.31& 0.19     &  1 &$<$0.65&0.19&1\\
Cr&Cr I &  --6.33 &    --0.13& 0.07&  3&    --0.65& 0.11&  3&     --0.63& 0.10&  3&     --0.59& 0.10&  3 &--0.55&0.17&3\\
Mn&Mn I &  --6.61 &    --0.33& 0.07&  3&    --1.05& 0.14&  3&     --1.10& 0.11&  3&     --0.79& 0.10&  3 &--0.92&0.21&2\\
Co&Co I &  --7.08 &      0.60& 0.19&  2&      0.49& 0.12&  4&       0.57& 0.10&  5&       0.59& 0.11&  5 &0.58&0.21&2\\
Ni&Ni I &  --5.75 &      0.24& 0.06&  4&    --0.12& 0.09&  5&       0.02& 0.10&  5&       0.18& 0.08&  5 &0.06&0.15&4\\
Sr&Sr II&  --9.10 & $<$--1.61& 0.11&  1\tablenotemark{f}&    --0.70& 0.16&  2&     --1.20& 0.16&  2&     --1.59& 0.15&  2 &0.10&0.23&2\\
Y &Y  II&  --9.76 &   $<$0.59& 0.15&  1\tablenotemark{g}& $<$--1.42& 0.15&  1\tablenotemark{h}&  $<$--0.46& 0.15&  1\tablenotemark{h}&  $<$--0.54& 0.15&  1\tablenotemark{h} &0.19&0.25&2\\
Ba &Ba II&  --9.87 & $<$--0.59& 0.14&  1&    --0.86& 0.23&  1&     --1.16& 0.22&  1&     --1.44& 0.21&  1 &--0.84&0.32&1\\
Eu &Eu II& --11.49 &   $<$1.56& 0.14&  1\tablenotemark{i}&   $<$0.82& 0.18&  1\tablenotemark{i}&    $<$0.75& 0.18&  1\tablenotemark{i}&    $<$0.68& 0.18&  1\tablenotemark{i} &$<$0.93&0.18&1\\
\enddata
\tablenotetext{a} {For Fe I the value tabulated under [X/Fe] is [Fe/H].}
\tablenotetext{b} {From Anders \& Grevesse (1989), except for C and N, which come from Lambert (1978).}
\tablenotetext{c} {For each star relative abundance [X/Fe], its 1$\sigma$ error, and number of lines, are given.}
\tablenotetext{d} {Al I 3961.5~{\AA}}
\tablenotetext{e} {Ti I 3998.6~{\AA}}
\tablenotetext{f} {Sr II 4077.7~{\AA}}
\tablenotetext{g} {Y II 3788.7~{\AA}}
\tablenotetext{h} {Y II 3774.3~{\AA}}
\tablenotetext{i} {Eu II 4129.7~{\AA}}
\end{deluxetable}
\end{center}

%TABLE6
\clearpage
\begin{center}
\tabletypesize{\scriptsize}
\begin{deluxetable}{lrrrrrr}
\tablecaption{CD--24$^{\rm o}$17504 \& CD~22949--037 ABUNDANCES FOR {\it BVRI} \& IRFM EFFECTIVE TEMPERATURES}
\tablehead{\multicolumn{1}{c}{} &\multicolumn{3}{c}{CD--24$^{\rm o}$17504} &\multicolumn{3}{c}{CS~22949--037}\\
\colhead{Species}  &{[X/Fe]$_{\rm This~work}$} &{[X/Fe]$_{\rm IRFM}$} &{$\Delta$[X/Fe]\tablenotemark{a}} &{[X/Fe]$_{\rm This~work}$} &{[X/Fe]$_{\rm IRFM}$} &{$\Delta$[X/Fe]\tablenotemark{a}}\\ 
{(1)}  &{(2)}     &{(3)}   &{(4)}    &{(5)} &{(6)} &{(7)}
}
\startdata
Fe I\tablenotemark{b} &    --3.37&    --3.15&   0.22&  --3.79&    --3.72&    0.07\\
Fe II                 &    --0.02&      0.00&   0.02&    0.00&      0.01&    0.01\\
Mg I                  &      0.47&      0.36& --0.11&    1.22&      1.17&  --0.05\\
Al I\tablenotemark{c} &    --0.79&    --0.80& --0.01&  --0.43&    --0.44&  --0.01\\
Si I                  &      0.04&      0.03& --0.01&    1.04&      0.99&    0.05\\
Ca I                  &      0.24&      0.18& --0.06&    0.45&      0.44&  --0.01\\
Sc II                 &      0.27&      0.37&   0.10&    0.13&      0.18&    0.05\\
Ti I\tablenotemark{d} &   $<$0.44&   $<$0.47&   0.03& $<$0.02&   $<$0.05&    0.03\\
Ti II                 &      0.27&      0.35&   0.08&    0.35&      0.37&    0.02\\
V  II                 &   $<$0.94&   $<$1.00&   0.06& $<$0.65&   $<$0.68&    0.03\\
Cr I                  &    --0.13&    --0.09&   0.04&  --0.55&    --0.51&    0.04\\
Mn I                  &    --0.33&    --0.27&   0.06&  --0.92&    --0.87&    0.05\\
Co I                  &      0.60&      0.65&   0.05&    0.58&      0.62&    0.04\\
Ni I                  &      0.24&      0.29&   0.05&    0.06&      0.08&    0.02\\
Sr II\tablenotemark{e}& $<$--1.61& $<$--1.48&   0.13&    0.10&      0.11&    0.01\\
Y  II\tablenotemark{f}&   $<$0.59&   $<$0.71&   0.12&    0.19&      0.24&    0.05\\
Ba II                 & $<$--0.59&   $<$--0.45& 0.14&  --0.84&    --0.77&    0.07\\
Eu II\tablenotemark{g}&   $<$1.56&   $<$1.69&   0.13& $<$0.93&   $<$0.99&    0.06\\
\enddata
\tablenotetext{a} {$\Delta$[X/Fe] = [X/Fe]$_{\rm IFRM}$--[X/Fe]$_{\rm This~work}$}
\tablenotetext{b} {For Fe I the value tabulated under [X/Fe] is [Fe/H].}
\tablenotetext{c} {Al I 3961.5~{\AA}}
\tablenotetext{d} {Ti I 3998.6~{\AA}}
\tablenotetext{e} {Sr II 4077.7~{\AA}}
\tablenotetext{f} {Y II 3774.3~{\AA}}
\tablenotetext{g} {Eu II 4129.7~{\AA}}
\end{deluxetable}
\end{center}

%TABLE7
\clearpage
\begin{center}
\tabletypesize{\scriptsize}
\begin{deluxetable}{lccccccr}
\tablecaption{COMPARISON OF TEMPERATURE CALIBRATIONS OF {\it BVRI} PHOTOMETRY}
\tablehead{
\colhead{    }&\multicolumn{5}{c}{IRFM scale}   & \\\\
\colhead{Star}& {$B-V$(3)}& {$B-V$(4)}& {$V-R$(5)}& {$R-I$(7)} &
$\langle IRFM\rangle$& {Table~4}&{$\Delta$T$_{\rm eff}$\tablenotemark{a}}\\
\nl
        {(1)} &{(2)} &{(3)} & {(4)} &{(5)}&{(6)} &{(7)} &{(8)} 
}
\startdata
CD--38$^\circ$245& 4709&4897&       &4891\tablenotemark{b} &4832& 4850&--18\\
CS 22172--002     & 4872&4905&       &4880\tablenotemark{b} &4886& 4900&--14\\
CS 22886--096     & 5157&5017&       &5145\tablenotemark{b} &5106& 5050&+56\\
CS 22949--037     & 4999&4976& 5035\tablenotemark{b} &      &5003& 4900&+103\\
\ \\
$\sigma_{\rm Alonso et al.}$         
                 &  167&  96&  150  & 150 & \nodata\\
\enddata
\tablenotetext{a} {$\Delta$T$_{\rm eff}$ = T$_{\rm eff}$(IRFM)--T$_{\rm eff}$(This work, Table~4)}
\tablenotetext{b} {Value is particularly uncertain.}
\end{deluxetable}
\end{center}

%TABLE8
\clearpage
\begin{center}
\tabletypesize{\scriptsize}
\begin{deluxetable}{lrrr}
\tablecaption{ROBUST SCATTER ESTIMATES FOR ELEMENTAL RATIOS IN METAL--POOR STARS}
\tablehead{
\colhead{[Fe/H] Range}  &{$\langle$[Fe/H]$\rangle$} &{N}   &{$S_{BI}$ (Errors on $S_{BI}$)}\\
           {(1)}        &{(2)}                      &{(3)} &{(4)}
}
\startdata
 & & & \\
$[$Mg/Fe$]$& & &\\          
 & & & \\
$\geq$ --0.5  & --0.18 &  39 &   0.12 (--0.01,+0.02)\\
--1.0 to --0.5 & --0.77 &  91 &   0.12 (--0.01,+0.01)\\
--1.5 to --1.0 & --1.23 &  42 &   0.15 (--0.01,+0.02)\\
--2.0 to --1.5 & --1.72 &  54 &   0.16 (--0.02,+0.02)\\
--2.5 to --2.0 & --2.24 &  27 &   0.16 (--0.02,+0.04)\\
$\leq$ --2.5  & --3.00 &  42 &   0.20 (--0.02,+0.04)\\
 & & & \\
$[$Al/Fe$]$& & &\\
 & & & \\
$\geq$ --0.5       & --0.16 &  34 &   0.14 (--0.01,+0.02)\\
--1.0 to --0.5 & --0.75 &  57 &   0.13 (--0.01,+0.02)\\
--1.5 to --1.0 & --1.23 &  21 &   0.20 (--0.03,+0.04)\\
--2.5 to --1.5 & --1.83 &  23 &   0.43 (--0.05,+0.07)\\
$\leq$ --2.5      & --3.06 &  33 &   0.21 (--0.03,+0.04)\\
 & & & \\
$[$Si/Fe$]$& & &\\
 & & & \\
$\geq$ --0.5       & --0.31 &  31 &   0.06 (--0.01,+0.02)\\
--1.0 to --0.5 & --0.77 &  94 &   0.11 (--0.01,+0.01)\\
--1.5 to --1.0 & --1.23 &  43 &   0.13 (--0.01,+0.01)\\
--2.0 to --1.5 & --1.71 &  52 &   0.13 (--0.01,+0.01)\\
--2.5 to --2.0 & --2.22 &  15 &   0.14 (--0.02,+0.04)\\
$\leq$ --2.5      & --3.05 &  29 &   0.33 (--0.04,+0.05)\\
 & & & \\
$[$Ca/Fe$]$& & &\\
 & & & \\
$\geq$ --0.5       & --0.18 &  39 &   0.01 (--0.01,+0.01)\\
--1.0 to --0.5 & --0.77 &  92 &   0.07 (--0.01,+0.01)\\
--1.5 to --1.0 & --1.23 &  41 &   0.08 (--0.01,+0.01)\\
--2.0 to --1.5 & --1.72 &  54 &   0.11 (--0.01,+0.02)\\
--2.5 to --2.0 & --2.23 &  28 &   0.09 (--0.02,+0.04)\\
$\leq$ --2.5      & --2.99 &  43 &   0.16 (--0.01,+0.02)\\
 & & & \\
$[$Sc/Fe$]$& & &\\
 & & & \\
$\geq$ --2.5       & --1.36 &  24  &  0.07 (--0.01,+0.01)\\
$\leq$ --2.5      & --3.04 &  32  &  0.19 (--0.02,+0.02)\\
 & & & \\
$[$Ti/Fe$]$& & &\\         
 & & & \\
$\geq$ --0.5       & --0.18 &  41 &   0.11 (--0.01,+0.02)\\
--1.0 to --0.5 & --0.77 &  91 &   0.09 (--0.01,+0.01)\\
--1.5 to --1.0 & --1.23 &  41 &   0.06 (--0.01,+0.01)\\
--2.0 to --1.5 & --1.73 &  52 &   0.12 (--0.01,+0.02)\\
--2.5 to --2.0 & --2.23 &  29 &   0.19 (--0.02,+0.04)\\
$\leq$ --2.5      & --3.01 &  41 &   0.19 (--0.02,+0.02)\\
 & & & \\
$[$Cr/Fe$]$& & &\\         
 & & & \\
$\geq$ --0.5       & --0.33 &  21  &  0.04 (--0.00,+0.01)\\ 
--1.0 to --0.5 & --0.77 &  87  &  0.04 (--0.00,+0.00)\\
--1.5 to --1.0 & --1.23 &  40  &  0.05 (--0.01,+0.01)\\
--2.0 to --1.5 & --1.72 &  46  &  0.08 (--0.02,+0.02)\\
--2.5 to --2.0 & --2.23 &  23  &  0.10 (--0.01,+0.02)\\
$\leq$ --2.5      & --2.99 &  40  &  0.20 (--0.02,+0.03)\\
 & & & \\
$[$Mn/Fe$]$& & &\\
 & & & \\
$\geq$ --2.5       & --1.14 &  26  &  0.06 (--0.01,+0.01)\\
$\leq$ --2.5      & --3.11 &  23  &  0.19 (--0.03,+0.04)\\
 & & & \\
$[$Co/Fe$]$& & &\\
 & & & \\
$\geq$ --2.5       & --1.11 &  16  &  0.07 (--0.01,+0.02)\\
$\leq$ --2.5      & --3.14 &  17  &  0.18 (--0.05,+0.07)\\
 & & & \\
$[$Ni/Fe$]$& & &\\         
 & & & \\
$\geq$ --0.5       & --0.19 &  42  &  0.06 (--0.01,+0.02)\\
--1.0 to --0.5 & --0.77 &  91  &  0.06 (--0.01,+0.01)\\
--1.5 to --1.0 & --1.23 &  42  &  0.08 (--0.01,+0.02)\\
--2.0 to --1.5 & --1.71 &  53  &  0.09 (--0.01,+0.01)\\
--2.5 to --2.0 & --2.22 &  20  &  0.12 (--0.02,+0.02)\\
$\leq$ --2.5      & --3.01 &  33  &  0.21 (--0.03,+0.01)\\
 & & & \\
$[$Sr/Fe$]$& & &\\               
 & & & \\
$\geq$ --2.5       & --1.50 &  33  &  0.26 (--0.03,+0.03)\\
$\leq$ --2.5      & --3.00 &  31  &  0.51 (--0.08,+0.01)\\
 & & & \\
$[$Ba/Fe$]$& & &\\
 & & & \\
$\geq$ --0.5       & --0.33 &  23  &  0.18 (--0.02,+0.07)\\
--1.0 to --0.5 & --0.77 &  89  &  0.10 (--0.01,+0.01)\\
--1.5 to --1.0 & --1.24 &  43  &  0.13 (--0.02,+0.03)\\
--2.0 to --1.5 & --1.72 &  52  &  0.20 (--0.03,+0.03)\\
--2.5 to --2.0 & --2.23 &  32  &  0.29 (--0.04,+0.07)\\
$\leq$ --2.5      & --2.97 &  25  &  0.43 (--0.05,+0.12)\\

\enddata
\end{deluxetable}
\end{center}

%TABLE9
\clearpage
\begin{center}
\begin{deluxetable}{lrr}
\tablecaption{PRODUCTION FACTORS FOR Z35B \&  CS~22949--037}
\tablehead{
\colhead{Element}  &{p(Z35B, Woosley \& Weaver)} &{p(CS~22949--037)} \\
           {(1)}        &{(2)}                      &{(3)} 
}
\startdata
C  & 4.17 & 2.17 \\
Mg & 3.34 & 3.34 \\
Al & 0.55 & 0.25 \\
Si & 0.09 & 2.03 \\
Ca & 0.00 & 0.29 \\
Sc & 0.00 & 0.03 \\
Ti & 0.00 & 0.12 \\
Cr & 0.00 & 0.01 \\
Mn & 0.00 & 0.00 \\
Fe & 0.00 & 0.00 \\
Co & 0.00 & 0.00 \\
Ni & 0.00 & 0.00 \\
\enddata
\end{deluxetable}
\end{center}

%FIGURE CAPTIONS
\newpage
\begin{figure} [h]  
FIGURE CAPTIONS  
%FIGURE1
\figurenum{1}
\caption {Spectra of the program stars in the G-band region, in comparison
with those of the more metal-rich archetype halo subgiant HD 140283 and giant
HD 122563. T$_{\rm eff}$/log~$g$/[Fe/H] values are also presented.}

%FIGURE2 
\figurenum{2} 
\caption {Comparison between the equivalent widths of the present work
(abscissa) and other investigations (ordinate).  (The diagonal lines
represent 1--1 relationships.)  For CD--24$^{\rm o}$17504 the literature
values are from Ryan et al. (1991) and are independent of the present work:
the data were obtained with a photon-counting detector, and equivalent widths
were measured by a different author.  For the other stars the comparison data
are from McWilliam et al. (1995a).  Values of the figure of merit F (\S 1) of
the data are indicated near the axes.}

%FIGURE3
\figurenum{3} 
\caption {Comparison of observed (thick lines) and synthetic
spectra (thin lines) in the region of violet CN at $\lambda$3883~{\AA} and CH
at $\lambda$4323~{\AA}.  On the right the three synthetic spectra differ in
steps of $\Delta$[C/Fe] = 0.3, while on the left the computations were made
with the best fit [C/Fe] determined at $\lambda$4323~{\AA} and steps of
$\Delta$[N/Fe] = 0.3. The adopted values of [C/Fe] and [N/Fe] are given in
parentheses in the right and left panels, respectively.}

%FIGURE4 
\figurenum{4} 
\caption {[C/Fe] vs [Fe/H] for metal-deficient stars. Filled circles come
from the present work; open circles from Ryan et al. (1991) and Norris et
al. (1997a); asterisks from McWilliam et al. (1995b), excluding values
designated uncertain; and crosses from Tomkin et al. (1992, 1995).  The 
dashed line represents the result of the Galactic chemical enrichment 
model of Timmes et al. (1995).}

%FIGURE5
\figurenum{5} 
\caption {[$\alpha$/Fe] vs. [Fe/H].  The left panels show the individual
data from this paper (filled circles); Ryan et al. (1991, 1996), Norris et
al. (2000) (open circles); Gratton (1989), Gratton et al. (1987, 1988, 1991,
1994) (asterisks); Nissen \& Schuster (1997) (open triangles); Carney et
al. (1997) (crosses); Stephens (1999) (open squares); and Fulbright (2000)
(plus signs). The right panels show the same data overlaid by five
curves. The central solid line represents the robust trend (the central loess
line) computed for the data as described in \S 4.1.2, flanked by lower and
upper loess lines.  The lower dashed curve represents the Galactic chemical
enrichment model of Timmes et al. (1995), while the upper one is their
calculation with the massive-star Fe yield reduced by a factor of two.}

%FIGURE6
\figurenum{6} 
\caption {[Al/Fe] vs. [Fe/H] (upper panels) and [Al/Mg] vs. [Fe/H] (lower
panels).  Symbols and curves are the same as in Figure~5, except that in the
lower right panel only the baseline Galactic chemical enrichment model of
Timmes et al. (1995) is shown since [Al/Mg] is barely affected by the Fe
yield.}

%FIGURE7
\figurenum{7} 
\caption {[Al/Mg] corrected for non-LTE effects, following Baum\"{u}ller \&
Gehren (1997) as described in the text, as a function of [Fe/H].  The dashed
curve is the baseline Galactic chemical enrichment result of Timmes et
al. (1995), while the thin lines are the lowess relations described in the
text.}

\end{figure}

\newpage
\begin{figure}  

%FIGURE8
\figurenum{8} 
\vspace{-60mm}
\caption {[Fe-peak/Fe] vs. [Fe/H]. Symbols and curves are the same as in Figure~5.}

%FIGURE9
\figurenum{9} 
\caption {[Neutron-capture element/Fe] vs. [Fe/H]. Symbols and curves are
the same as in Figure~5, except that data from Gilroy et al. (1988) have also
been included, as diamonds.} 

%FIGURE10
\figurenum{10} 
\caption {The robust scatter parameter $S_{BI}$ as a function of [Fe/H] for 12
relative abundances.  See Table~8 for data, and text for discussion.}

%FIGURE11
\figurenum{11} 
\caption {(a) Relative abundances of CS~22949--037, relative to the average
values of the other four stars of the present sample, as a function of atomic
species. (b) Relative abundances of CS~22885--096, relative to the average
values of CD--24$^{\rm o}$17504, CD--38$^{\rm o}$245, and CS~22172--002, as a
function of atomic species.}

%FIGURE12 
\figurenum{12} 
\caption {Logarithm of the ratio of the masses of element X and Mg,
log(M$_{\rm X}$/M$_{\rm Mg}$), as a function of atomic species.  The continuous
lines represent the results for the zero-heavy-element supernova models Z35B
(lower) and Z35C (upper) of Woosley \& Weaver (1995), while the closed and
open stars show the corresponding values for CS~22949--037 and the averages
of those for the other four stars observed in the present work,
respectively.}

\end{figure}

%FIGURES

\newpage
%FIGURE1
\begin{figure} [h]  
\plotone{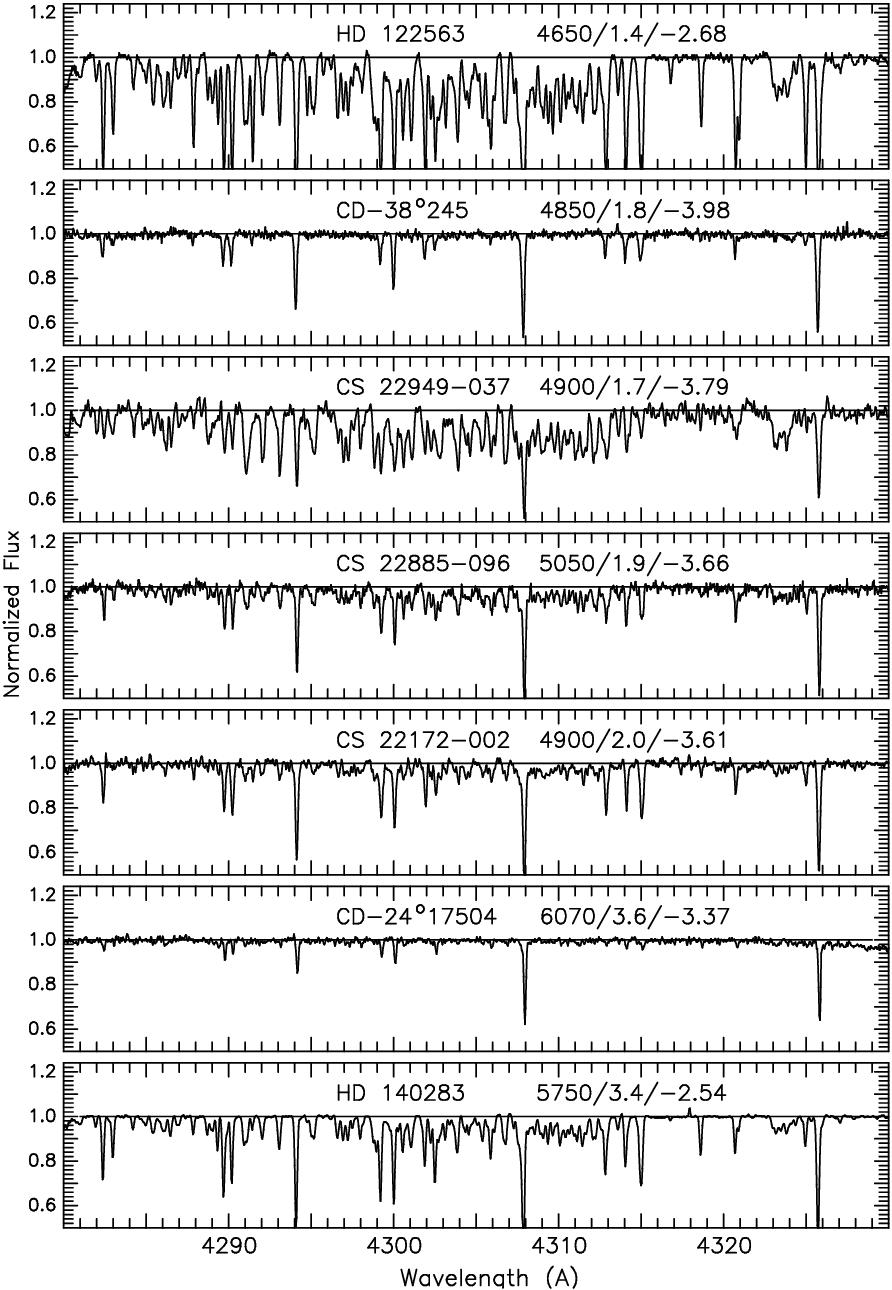}
\caption {}
\end{figure}

\newpage
%FIGURE2 
\begin{figure} [h]  
\plotone{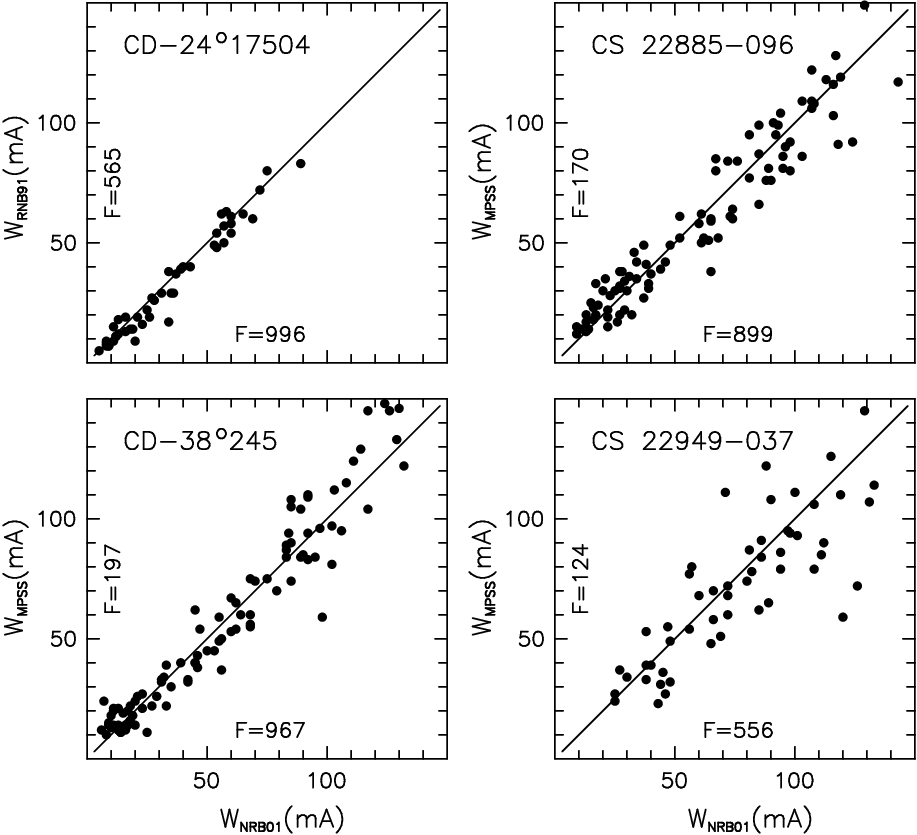}
\caption {}
\end{figure}

\newpage
%FIGURE3
\begin{figure} [h]  
\plotone{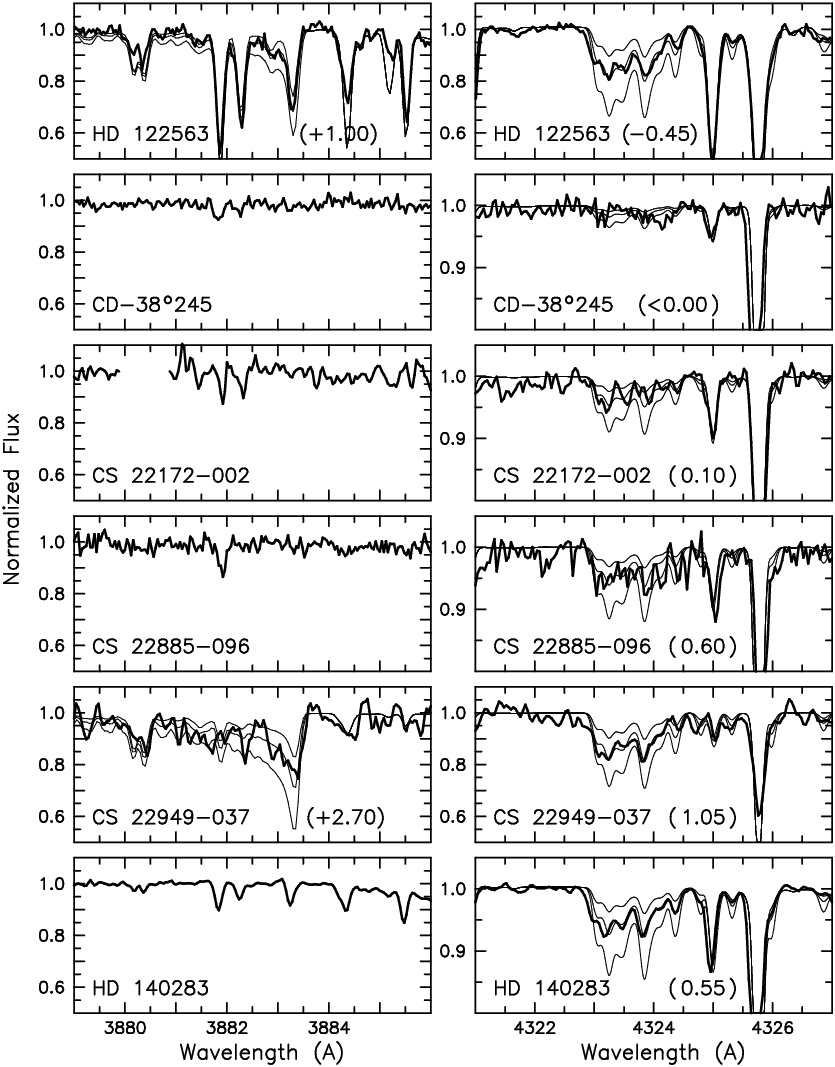}
\caption {}
\end{figure}

\newpage
%FIGURE4 
\begin{figure} [h]  
\plotone{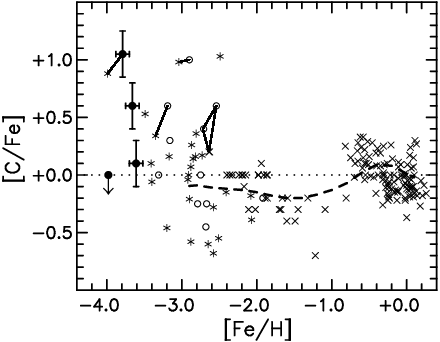}
\caption {}
\end{figure}

\newpage
%FIGURE5
\begin{figure} [h]  
\plotone{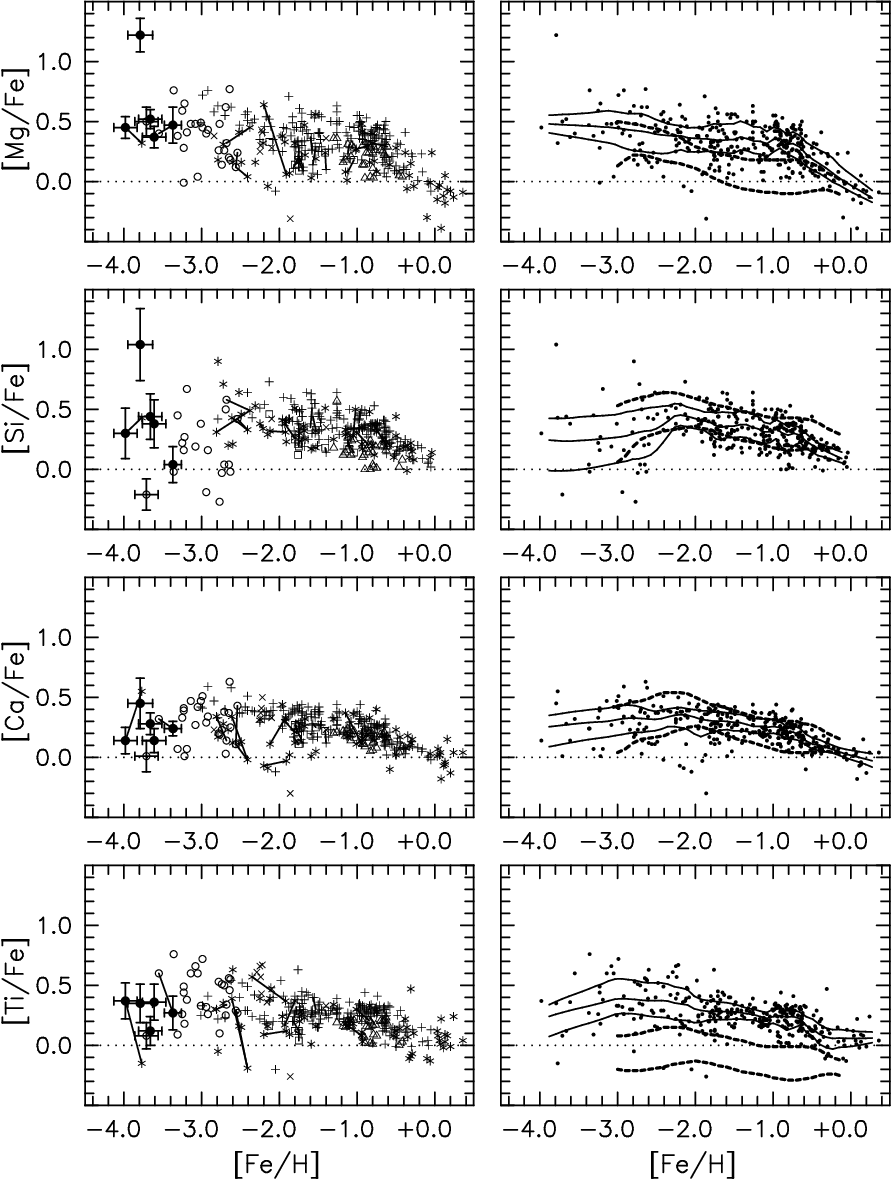}
\caption {}
\end{figure}

\newpage
%FIGURE6
\begin{figure} [h]  
\plotone{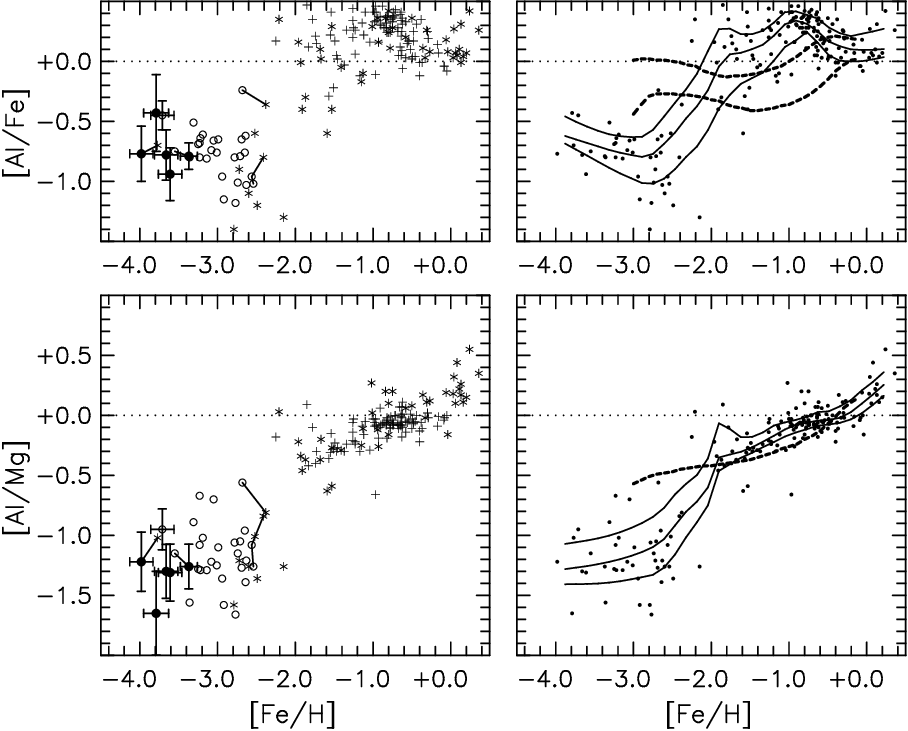}
\caption {}
\end{figure}

\newpage
%FIGURE7
\begin{figure} [h]  
\plotone{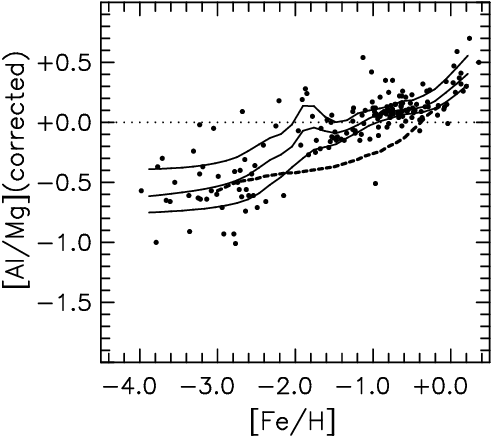}
\caption {}
\end{figure}

\newpage
%FIGURE8
\begin{figure} [h]  
\plotone{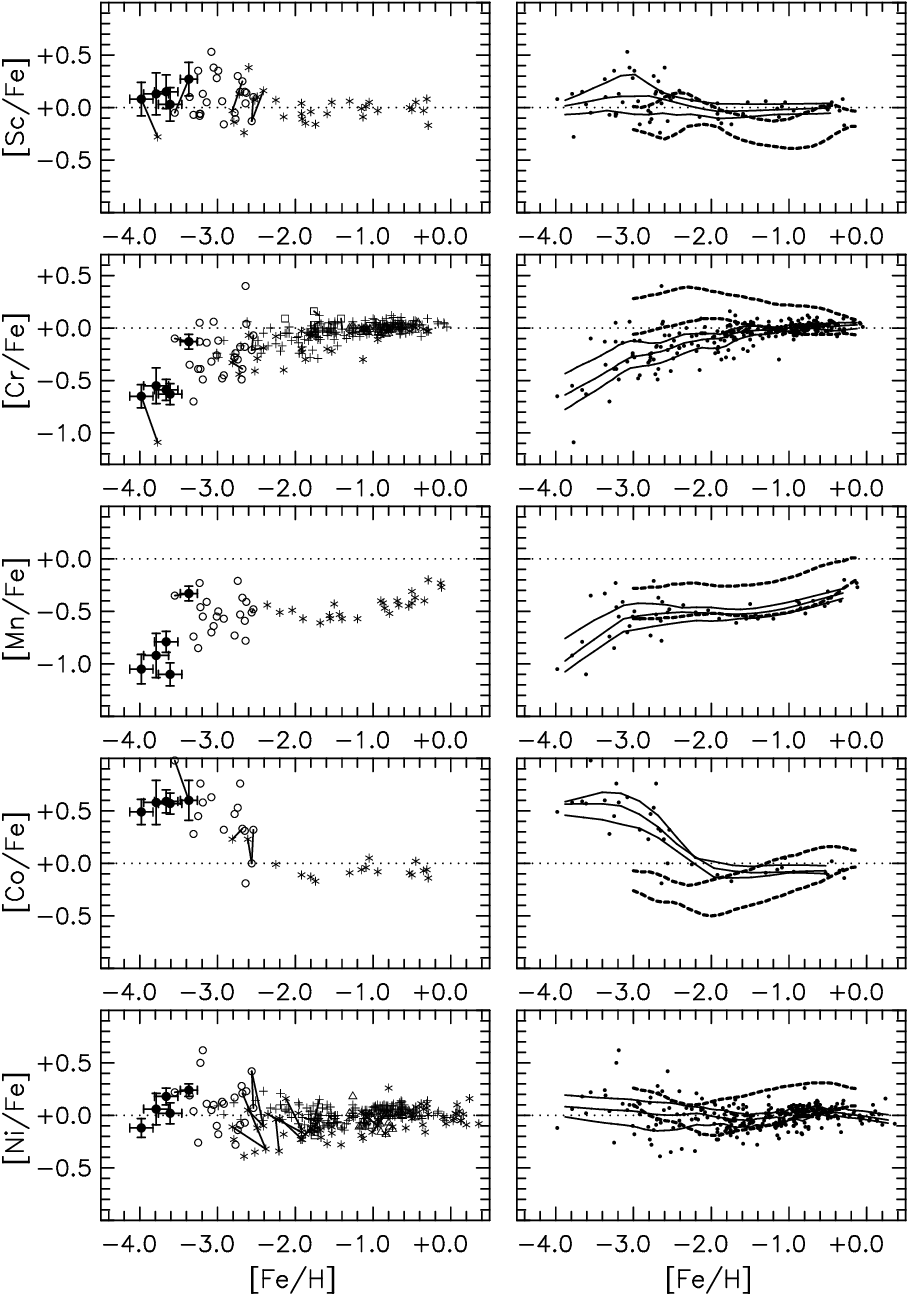}
\caption {}
\end{figure}

\newpage
%FIGURE9
\begin{figure} [h]  
\plotone{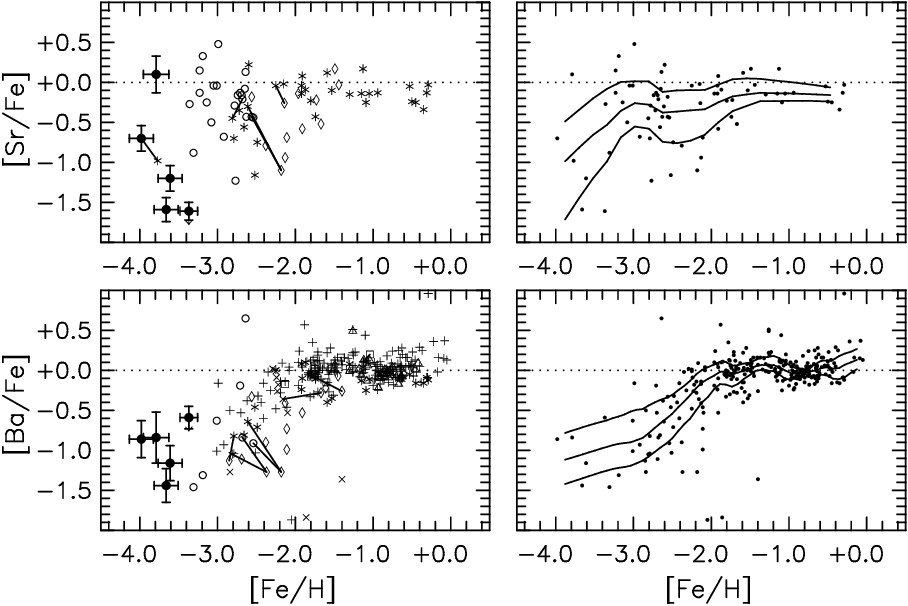}
\caption {} 
\end{figure}

\newpage
%FIGURE10
\begin{figure} [h]  
%\plotone{f10.eps}
\centering \leavevmode 
\epsfxsize = 8.3cm
\epsfbox [80 100 300 670] {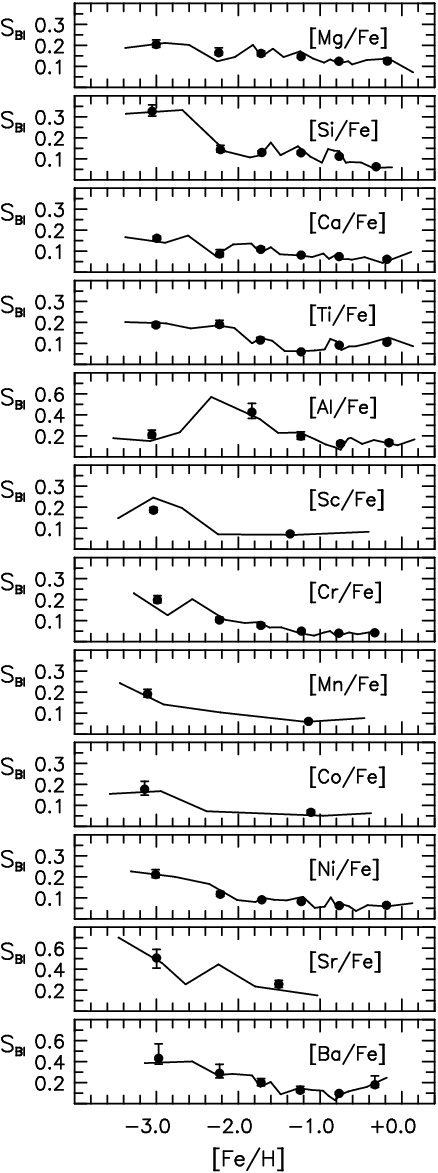}
\caption {}
\end{figure}

\newpage
%FIGURE11
\begin{figure} [h]  
\plotone{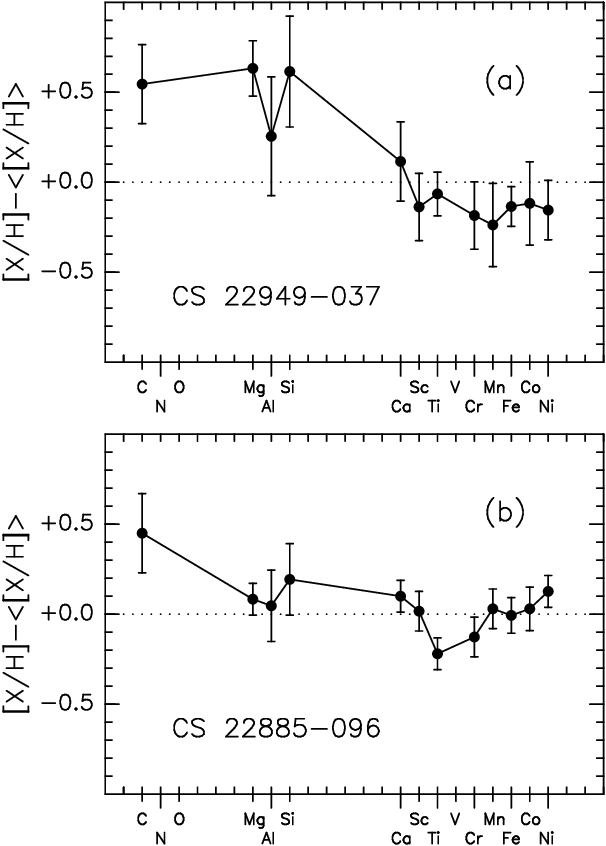}
\caption {}
\end{figure}

\newpage
%FIGURE12 
\begin{figure} [h]  
\plotone{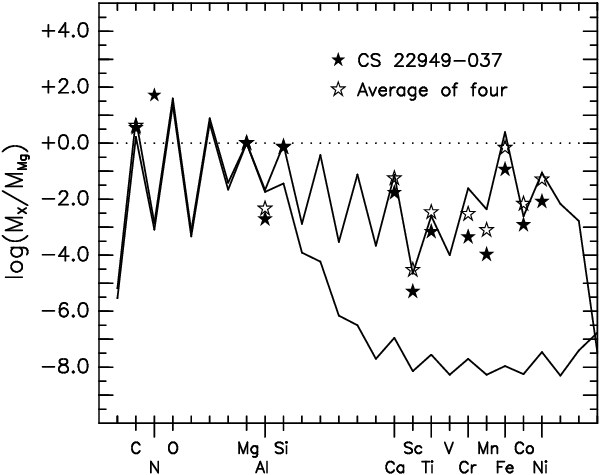}
\caption {}
\end{figure}

\end{document}